\documentclass[12pt, article, onecolumn]{IEEEtran}
\usepackage{amsmath}
\usepackage{amssymb}
\usepackage{mathrsfs}
\usepackage{cite}
\usepackage{epsfig}
\usepackage{epsf}
\usepackage{theorem}
\usepackage{graphics}

\newtheorem{thm}{Theorem}
\newtheorem{lem}{Lemma}

\begin{document}
\title{On the Throughput Maximization in Dencentralized Wireless Networks\\
\thanks{$^{*}$ This work is financially supported by Nortel Networks and the corresponding matching funds by the Natural Sciences
and Engineering Research Council of Canada (NSERC), and Ontario Centers of Excellence (OCE).}
\thanks{$^{*}$ The material in this paper was presented in part at the IEEE
International Symposium on Information Theory (ISIT), Nice, France,
June 24-29, 2007 \cite{JamshidISIT1} and the 41th Conference on IEEE
Information Sciences and Systems (CISS), Johns Hopkins University,
Baltimore, MD, March 14-16, 2007 \cite{JamshidCISS07}.}}
%, and
%reported in the technical report number UW-ECE \#2007-14, April
%2007, the technical report number UW-ECE \#2007-20 August 2007, and
%the technical report number UW-ECE \#2007-33 November 2007,
%University of Waterloo, Waterloo, Ontario, Canada.}}

\author{ Jamshid Abouei, Alireza Bayesteh, Masoud Ebrahimi, and Amir K. Khandani \\
\small Coding and Signal Transmission Laboratory (www.cst.uwaterloo.ca)\\
Department of Electrical and Computer Engineering, University of Waterloo\\
Waterloo, Ontario, Canada, N2L 3G1 \\
Tel: 519-884-8552, Fax: 519-888-4338\\
Emails: \{jabouei, alireza, masoud, khandani\}@cst.uwaterloo.ca}

\maketitle

%\markboth{To be Submitted to IEEE Transactions on Information Theory}{}

\begin{abstract}
A distributed single-hop wireless network with $K$ links is considered, where the links are 
partitioned into a fixed number ($M$) of clusters each operating in a subchannel with 
bandwidth $\frac{W}{M}$. The subchannels are assumed to be orthogonal to each other. A general shadow-fading model, described by parameters $(\alpha,\varpi)$, is considered where 
$\alpha$ denotes the probability of shadowing and $\varpi$ ($\varpi \leq 1$)  represents 
the average cross-link gains. The main goal of this paper is to find the maximum  network
throughput in the asymptotic regime of $K \to \infty$, which is achieved by: i)~proposing 
a distributed and non-iterative power allocation strategy, where the objective of each user 
is to maximize its best estimate (based on its local information, i.e., direct channel gain) 
of the average network throughput, and ii)~choosing the optimum value for $M$. In the first 
part of the paper, the network throughput is defined as the \textit{average sum-rate} of the 
network, which is shown to scale as $\Theta (\log K)$. Moreover,  it is proved that in the 
strong interference scenario, the optimum power allocation strategy for each user is a 
threshold-based on-off scheme. In the second part, the network throughput is defined as the 
\textit{guaranteed sum-rate}, when the outage probability approaches zero. In this scenario, 
it is demonstrated that the on-off power allocation scheme maximizes the throughput, which 
scales as $\frac{W}{\alpha \varpi} \log K$.  Moreover, the optimum spectrum sharing for 
maximizing the average sum-rate and the guaranteed sum-rate  is achieved at $M=1$.
\end{abstract}

\begin{center}
\end{center}

\section{Introduction}

\subsection{History}
A primary challenge in wireless networks is to use available
resources efficiently so that the network throughput is maximized.
Throughput maximization in multi-user wireless networks has been
addressed from different perspectives; resource allocation
\cite{LiangITIT1007, KumaranITWC0505, HollidayISIT2004}, scheduling
\cite{ViswanathITIT0602}, routing by using relay nodes
\cite{YehBerryITIT1007}, exploiting mobility of the nodes
\cite{GrossglauserIACM0802} and exploiting channel characteristics
(e.g., power decay-versus-distance law \cite{GuptaITIT2000,
KulkaraniITIT0604, XieITIT0504}, geometric pathloss and fading
\cite{JovicicViswanathITIT1104, XueXieKumarITIT0305, NiesenITIT2007}).

Among different resource allocation strategies, power and spectrum
allocation have long been regarded as efficient tools to mitigate
the interference and improve the network throughput. In recent
years, power and spectrum allocation schemes have been extensively
studied in cellular and multihop wireless networks
\cite{SampathPIMRC95, KumaranITWC0505, ElBattITWC0104,
WassermanITWC0603, HanITC0805, KatzelaIPC0696, KianiISIT2006,
LiangITIT1007}. In \cite{KatzelaIPC0696}, the authors provide a
comprehensive survey in the area of resource allocation, in
particular in the context of spectrum assignment. Much of these
works rely on centralized and cooperative algorithms. Clearly,
centralized resource allocation schemes provide a significant
improvement in the network throughput over decentralized
(distributed) approaches. However, they require extensive knowledge
of the network configuration. In particular, when the number of
nodes is large, deploying such centralized schemes may not be
practically feasible.
%In addition, in a cooperative wireless
%network, when the number of nodes becomes large, the amount of
%exchanged control information grows extensively. This is
%particularly problematic for time-varying networks, as the
%algorithms can not track the channel variations.
Due to significant challenges in using centralized approaches, the
attention of the researchers has been drawn to the decentralized
resource allocation schemes \cite{YatesJSAC0995, FoschiniITVT1193,
SaraydarITC0202, HuangJSAC2006, EtkinTse2007, JindalAllerton07}.

In decentralized schemes, the decisions concerning network
parameters (e.g., rate and/or power) are made by the individual
nodes based on their local information. The local decision
parameters that can be used for adjusting the rate are the
Signal-to-Interference-plus-Noise Ratio (SINR) and the direct
channel gain. Most of the works on decentralized throughput
maximization target the SINR parameter by using iterative algorithms
\cite{SaraydarITC0202, HuangJSAC2006, EtkinTse2007}. This leads to
the use of game theory concepts~\cite{Osbornebook2004} where the
main challenge is the convergence issue. For instance, Etkin
\textit{et al.} \cite{EtkinTse2007} develop power and spectrum
allocation strategies by using game theory. Under the assumptions of
the omniscient nodes and strong interference, the authors show that
Frequency-Division Multiplexing (FDM) is the optimal scheme in the
sense of throughput maximization. They use an iterative algorithm
that converges to the optimum power values. In \cite{HuangJSAC2006}, 
Huang \textit{et al.} propose an iterative power control algorithm 
in an ad hoc wireless network, in which receivers broadcast 
adjacent channel gains and interference prices to optimize the 
network throughput. However, this algorithm incurs a great amount
of overhead in large wireless networks. 

A more practical approach is to rely on the channel gains as local
decision parameters and avoid iterative schemes. Motivated by this
consideration, we study the throughput maximization of a distributed
wireless network with $K$ links, operating in a bandwidth of $W$. To
mitigate the interference, the links are partitioned into a fixed 
number ($M$) of clusters, each operating in a subchannel with bandwidth
$\frac{W}{M}$, where the subchannels are orthogonal to each other.
Throughput maximization of the underlying network is achieved by
proposing a distributed and non-iterative power allocation strategy
based on the direct channel gains, and then choosing the optimum
value for $M$.

\subsection{Contributions and Relations to Previous Works}
In this paper, we study the throughput maximization of a spatially
distributed wireless network with $K$ links, where the sources and
their corresponding destinations communicate directly with each
other without using relay nodes. Wireless networks using unlicensed
spectrum (e.g. Wi-Fi systems based on IEEE 802.11b standard
\cite{Ohrtman2003}) are a typical example of such networks.  The
cross-link channel gains are assumed to be Rayleigh-distributed with
shadow-fading, described by parameters $(\alpha,\varpi)$, where
$\alpha$ denotes the probability of shadowing and $\varpi$ ($\varpi \leq 1$)
represents the statistical average of the Rayleigh distribution.

The above configuration differs from the geometric models proposed
in \cite{GuptaITIT2000, GrossglauserIACM0802, KulkaraniITIT0604,
XieITIT0504, JindalITWC2007}, in which the signal power decays based
on the distance between nodes. Unlike~\cite{FoschiniITVT1193,
SaraydarITC0202, HuangJSAC2006, EtkinTse2007} which relies on an
iterative algorithm using SINR, we assume that each transmitter
adjusts its power solely based on its direct channel gain.

If each user maximizes its rate selfishly, the optimum power
allocation strategy for all users is to transmit with full power.
This strategy results in excessive interference, degrading the
average network throughput. To prevent this undesirable effect, one
should consider the negative impact of each user's power on other
links. A reasonable approach for each user is to choose a
non-iterative power allocation strategy to maximize its best local
estimate of the network throughput.

The network throughput in this paper is defined in two ways: i)~\textit{average 
sum-rate} and ii)~\textit{guaranteed sum-rate}. It is established that the average 
sum-rate in the network scales at most as $\Theta (\log K)$ in the asymptotic case of 
$K \to \infty$. This order is achievable by the distributed \textit{threshold-based 
on-off scheme} (i.e., links with a direct channel gain above certain threshold 
transmit at full power and the rest remain silent). Moreover, in the strong 
interference scenario, the on-off power allocation scheme is the optimal strategy. 
In addition, the on-off power allocation scheme is always optimal for maximizing 
the guaranteed sum-rate in the network, which is shown to scale as $\frac{W}{\alpha \varpi} \log K$.
These results are  different from the result of~\cite{Masoud2006}
where the authors use a similar on-off scheme for $M=1$ and prove
its optimality only among all \textit{on-off schemes}. This work
also differs from \cite{HassibiITIT0706} and \cite{Bettstetter2005}
in terms of the network model. We use a distributed power allocation
strategy in a single-hop network, while \cite{HassibiITIT0706} and
\cite{Bettstetter2005} consider an ad hoc network model with random
connections and relay nodes.

We optimize the average network throughput in terms of the number of
the clusters, $M$. It is proved that the maximum average sum-rate and 
the guaranteed sum-rate  of the network for every value of $\alpha$ and 
$\varpi$ is achieved at $M=1$. In other words, splitting the bandwidth 
$W$ into $M$ orthogonal sub-channels does not increase the throughput.

The rest of the paper is organized as follows. In Section \ref{model}, 
the network model and objectives are described. The distributed on-off 
power allocation strategy and the network average sum-rate are presented 
in Section \ref{power}. We analyze the network guaranteed sum-rate in Section
\ref{guaranteed}. Finally, in Section \ref{conclusion1}, an overview
of the results and some conclusion remarks are presented.

\subsection{Notations}
For any functions $f(n)$ and $g(n)$ \cite{KnuthACM67}:
\begin{itemize}
\item [$\bullet$] $f(n)=O(g(n))$ means that $\lim_{n \to \infty} \Big \vert \frac{f(n)}{g(n)} \Big \vert < \infty$.
\item [$\bullet$] $f(n)=\mathit{o}(g(n))$ means that $\lim_{n \to \infty} \Big \vert \frac{f(n)}{g(n)} \Big \vert =0$.
\item [$\bullet$] $f(n)=\omega(g(n))$ means that $\lim_{n \to \infty} \frac{f(n)}{g(n)} = \infty$.
\item [$\bullet$] $f(n)=\Omega(g(n))$ means that $\lim_{n \to \infty}  \frac{f(n)}{g(n)}  > 0$.
\item [$\bullet$] $f(n)=\Theta (g(n))$ means that $\lim_{n \to \infty} \frac{f(n)}{g(n)}=c$, where $0<c<\infty$.
\item [$\bullet$] $f(n) \sim g(n)$ means that $\lim_{n \to \infty} \frac{f(n)}{g(n)}=1$.
\item [$\bullet$] $f(n) \lesssim g(n)$ means that $\lim_{n \to \infty} \frac{f(n)}{g(n)} \leq 1$.
\item [$\bullet$] $f(n) \approx g(n)$ means that $f(n)$ is approximately equal to $g(n)$, i.e., if 
we replace $f(n)$ by $g(n)$ in the equations, the results still hold.
\end{itemize}

Throughout the paper, we use $\log(.)$ as the natural logarithm function and $\mathbb{P} \{.\}$
denotes the probability of the given event. Boldface letters denote vectors; and for a random variable $x$,
$\bar{x}$ means $\mathbb{E}[x]$, where $\mathbb{E}[.]$ represents the expectation operator. $\mbox{R\!H(.)}$ 
represents the right hand side of the equations. 

\section{Network Model and Objectives}\label{model}
\subsection{Network Model}
In this work, we consider a single-hop wireless network consisting
of $K$ pairs of nodes\footnote{The term ``pair" is used to describe
a transmitter and its corresponding receiver, while the term ``user"
is used only for the transmitter.} indexed by $\lbrace 1,...,K
\rbrace$, operating in bandwidth $W$. All the nodes in the network
are assumed to have a single antenna. The links are assumed to be
randomly divided into $M$ clusters denoted by $\mathbb{C}_{j}$,
$j=1,...,M$ such that the number of links in all clusters are the
same. Without loss of generality, we assume that $\mathbb{C}_{j}
\triangleq \lbrace (j-1)n+1,...,jn\rbrace $, where $n \triangleq
\frac{K}{M}$ denotes the cardinality of the set $\mathbb{C}_{j}$
which is assumed to be known to all users\footnote{It is assumed
that $K$ is divisible by $M$, and hence, $n=\frac{K}{M}$ is an
integer number.}. To eliminate the mutual interference among the
clusters, we assume an $M$-dimensional orthogonal coordinate system in which
the bandwidth $W$ is split into $M$ disjoint subchannels each with
bandwidth $\frac{W}{M}$. It is assumed that the links in
$\mathbb{C}_{j}$ operate in subchannel $j$. We also assume that $M$ is fixed, i.e., it does not scale with $K$. The power of Additive
White Gaussian Noise (AWGN) at each receiver is $\frac{N_{0}W}{M}$,
where $N_{0}$ is the noise power spectral density.

The channel model is assumed to be flat Rayleigh fading with the shadowing effect.
The channel gain\footnote{In this paper, \textit{channel gain} is
defined as the square magnitude of the \textit{channel
coefficient}.} between transmitter $k$ and receiver $i$ is
represented by the random variable $\mathcal{L}_{ki}$. For $k=i$,
the \textit{direct channel gain} is defined as
$\mathcal{L}_{ki} \triangleq h_{ii}$ where $h_{ii}$ is exponentially
distributed with unit mean (and unit variance). For $k \neq i$, the
\textit{cross channel gains} are defined based on a shadowing model as
follows:
\begin{eqnarray}\label{eqn: 01}
    \mathcal{L}_{ki} \triangleq
\left\{\begin{array}{ll}
\beta_{ki} h_{ki}   ,    & \textrm{with probability}~~~~\alpha \\
0 ,  &  \textrm{with probability}~~ 1-\alpha,
\end{array} \right.
\end{eqnarray}
where $h_{ki}$'s have the same distribution as $h_{ii}$'s, $0 \leq
\alpha \leq 1$ is a fixed parameter, and the random variable
$\beta_{ki}$, referred to as the \textit{shadowing factor},
is independent of $h_{ki}$ and satisfies the following conditions:
\begin{itemize}
\item $\beta_{\min} \leq \beta_{ki} \leq \beta_{\max}$, where $\beta_{\min} >0$ and $\beta_{\max}$ is finite,
\item $\mathbb{E} \big[\beta_{ki}\big] \triangleq \varpi \leq 1$.
\end{itemize}
It is also assumed that $\{\mathcal{L}_{ki}\}$ and $\{\beta_{ki}\}$ are mutually independent random variables for different $(k,i)$.
 
All the channels in the
network are assumed to be quasi-static block fading, i.e., the
channel gains remain constant during one block and change
independently from block to block. In addition, we assume that each
transmitter knows its direct channel gain.

We assume a homogeneous network in the sense that all the links have
the same configuration and use the same protocol.  We denote the
transmit power of user $i$ by $p_{i}$, where $p_{i}\in
\mathscr{P}\triangleq [0,\textrm{P}_{max}]$. The vector
$\textbf{\textrm{P}}^{(j)}=(p_{(j-1)n+1},...,p_{jn})$ represents the
power vector of the users in $\mathbb{C}_{j}$. Also,
$\textbf{\textrm{P}}^{(j)}_{-i}$ denotes the vector consisting of
elements of $\textbf{\textrm{P}}^{(j)}$ other than the $i^{th}$
element, $i \in \mathbb{C}_{j}$. To simplify the notations, we
assume that the noise power $\frac{N_{0}W}{M}$ is normalized by
$\textrm{P}_{max}$. Therefore, without loss of generality, we assume that $\textrm{P}_{max}=1$. Assuming that the transmitted signals are
Gaussian, the interference term seen by link $i \in \mathbb{C}_{j}$
will be Gaussian with power
\begin{equation}\label{interf1}
I_{i}=\sum_{\substack{k  \in \mathbb{C}_{j} \\ k\neq i}} \mathcal{L}_{ki}p_{k}.
\end{equation}
Due to the orthogonality of the allocated sub-channels, no
interference is imposed from links in $\mathbb{C}_{k}$ on links in
$\mathbb{C}_{j}$, $k\neq j$. Under these assumptions, the achievable
data rate of each link $i \in \mathbb{C}_{j}$ is expressed as
\begin{equation}\label{eqn: 02}
R_{i}(\textbf{\textrm{P}}^{(j)},\boldsymbol{\mathcal{L}}_{i}^{(j)})=\dfrac{W}{M}\log \left( 1+  \dfrac{h_{ii}p_{i}}{I_{i}+\frac{N_{0}W}{M}} \right),
\end{equation}
where $\boldsymbol{\mathcal{L}}_{i}^{(j)} \triangleq
(\mathcal{L}_{((j-1)n+1)i},...,\mathcal{L}_{(jn)i})$. To analyze the
performance of the underlying network, we use the following performance metrics:
\begin{itemize}
\item \textbf{Network Average Sum-Rate:} 
\begin{equation} \label{eqn: 05}
%\bar{R}_{ave} \triangleq \sum_{j=1}^{M} \bar{R}_{ave}^{(j)},
\bar{R}_{ave} \triangleq \mathbb{E}\left [\sum_{j=1}^{M}\sum_{l \in \mathbb{C}_{j}} R_{l}(\textbf{\textrm{P}}^{(j)},\boldsymbol{\mathcal{L}}_{l}^{(j)}) \right],
\end{equation}
where the expectation is computed with respect to $\boldsymbol{\mathcal{L}}_{l}^{(j)}$. This metric is used when there is no decoding delay constraint, i.e., decoding is performed over arbitrarily large number of blocks.
\item \textbf{Network Guaranteed Sum-Rate:} 
\begin{equation} \label{gau}
\bar{R}_{g} \triangleq  \sum_{j=1}^{M}\sum_{l \in \mathbb{C}_{j}} \mathbb{E}_{h_{_{ll}}}  \left[ R^{*}(h_{ll}) \right],
\end{equation}
 in which for all $h_{ll}$, $l \in \mathbb{C}_j$, we have
\begin{equation}
 R^{*}(h_{ll}) \triangleq  \sup \, \, {R(h_{ll})},
\end{equation}
such that
\begin{equation} \label{gau2}
 \mathbb{P} \left \{ R_{l}(\textbf{\textrm{P}}^{(j)},\boldsymbol{\mathcal{L}}_{l}^{(j)}) <  R(h_{ll}) \right \} \rightarrow 0.
\end{equation}
This metric is useful when there exists a stringent decoding delay constraint, i.e, decoding must be performed over each separate block, and a single-layer code is used. In this case, as the transmitter does not have any information about the interference term, an outage event may occur. Network guaranteed throughput is the average sum-rate of the network which is guaranteed for all channel realizations.
\end{itemize}

\subsection{Objectives}
%Depending on the value of $\mathbb{E}[I_{i}]$, we investigate the following cases:

%\begin{itemize}
%\item \textbf{Strong Interference ($\mathbb{E}[I_{i}]=\omega(1)$):} For this case, we first prove that the optimum power allocation 
%strategy that maximizes the network throughput is the threshold based on-off power allocation policy. under 
%this scheme, we then obtain the network throughput for every value of $M$ and show that the 
%maximum throughput is obtained at $M=1$. 
%\item \textbf{Moderate Interference ($\mathbb{E}[I_{i}]=\Theta(1)$):} For this case, we first obtain an upper bound and a lower bound 
%for the network throughput. Then, we prove that the maximum throughput is achieved at $M=1$. 
%\item \textbf{Weak Interference ($\mathbb{E}[I_{i}]=o(1)$):} For this case, we prove that the maximum achievable throughput is 
%obtained at $M=1$.
%\end{itemize}

\textbf{Part I: Maximizing the network average sum-rate:} The main objective of the first 
part of this paper is to maximize the network average sum-rate when the interference is 
strong enough, i.e., $\mathbb{E}[I_{i}]=\omega(1)$. This is achieved by:
\begin{itemize}
\item [-] Proposing a distributed and non-iterative power allocation strategy, where each user maximizes its
best estimate (based on its local information, i.e., direct channel gain) of the average network sum-rate.
\item [-] Choosing the optimum value for $M$.
\end{itemize}

To address this problem, we first define a utility function for link $i \in \mathbb{C}_{j}$ ($j=1,...,M$) 
that describes the average sum-rate of the links in cluster $\mathbb{C}_{j}$ as follows
\begin{equation}\label{utility1}
u_{i}(p_{i},h_{ii}) \triangleq \mathbb{E} \left[\sum_{l \in \mathbb{C}_{j}}R_{l}(\textbf{\textrm{P}}^{(j)},\boldsymbol{\mathcal{L}}_{l}^{(j)}) \right],
\end{equation}
where the expectation is computed with respect to
 $\lbrace \mathcal{L}_{kl}
\rbrace_{k,l\in \mathbb{C}_{j}}$ excluding $k=l=i$ (namely
$h_{ii}$). As mentioned earlier, $h_{ii}$ is considered as the local
(known) information for link $i$, however, all the other gains are
unknown to user $i$ which is the reason behind statistical averaging
over these parameters in (\ref{utility1}). User $i$ selects its
power using
\begin{equation}\label{DPA}
\hat{p}_{i}=\textrm{arg}~\max_{p_{i}\in \mathscr{P}}~u_{i}(p_{i},h_{ii}).
\end{equation}
It will be shown that when the number of links is large and the interference is strong enough, 
the optimum power allocation strategy for the optimization problem in (\ref{DPA}) is the 
on-off power scheme. Assuming that the channel gains change independently from block to block, 
each user updates its on-off decision based on its direct channel gain in each block.
Given the optimum power vector
$\hat{\textbf{\textrm{P}}}^{(j)}=(\hat{p}_{(j-1)n+1},...,\hat{p}_{jn})$
obtained from (\ref{DPA}), the network average sum-rate is then computed as (\ref{eqn: 05}). Next, we choose the optimum
value of $M$ such that the network average sum-rate is maximized, i.e.,
\begin{equation} \label{eqn: 5}
\hat{M}=\textrm{arg}~\max_{M}~\bar{R}_{ave}.
\end{equation}
Also, for the moderate and the weak interference regimes (i.e., $\mathbb{E}[I_{i}]=O(1)$), 
we obtain upper bounds for the network average sum-rate.

\textbf{Part II: Maximizing the network guaranteed sum-rate:} The main objective of the 
second part is finding the maximum achievable network guaranteed sum-rate in the asymptotic 
case of $K \to \infty$. For this purpose, a lower bound and an upper-bound on the network 
guaranteed sum-rate are presented and shown to converge to each other as $K \to \infty$. 
Also, the optimum value of $M$ is obtained.

\section{Network Average Sum-rate}\label{power}
\subsection{Strong Interference Scenario ($\mathbb{E}[I_i]=\omega (1)$)}
In order to maximize the average sum-rate of the network, we first find the optimum power allocation policy.
Using (\ref{utility1}), we can express the utility function of link
$i \in \mathbb{C}_{j},~~j=1,...,M,$ as
\begin{equation}\label{DPAA}
u_{i}(p_{i},h_{ii})= \bar{R}_{i}(p_{i},h_{ii})+\sum_{\substack{l \in
\mathbb{C}_{j} \\ l \neq i}}\bar{R}_{l}(p_{i}),
\end{equation}
where
\begin{equation}\label{term1}
\bar{R}_{i}(p_{i},h_{ii})=\mathbb{E}\left[ \dfrac{W}{M} \log \left( 1+\dfrac{h_{ii}p_{i}}{I_{i}+\frac{N_{0}W}{M}}\right) \right],
\end{equation}
with the expectation computed with respect to $I_{i}$ defined in
(\ref{interf1}), and
\begin{eqnarray}
\bar{R}_{l}(p_{i}) &=& \mathbb{E} \left[R_{l}(\textbf{\textrm{P}}^{(j)},\boldsymbol{\mathcal{L}}_{l}^{(j)}) \right]\\
&=& \mathbb{E}\left[\dfrac{W}{M} \log \left(1+\dfrac{h_{ll}p_{l}}{I_{l}+\frac{N_{0}W}{M}}\right)\right]\\
\label{term2} &=& \mathbb{E}\left[\dfrac{W}{M} \log \left( 1+
\dfrac{h_{ll}p_{l}}{\mathcal{L}_{il}p_{i}+\sum_{k\neq l,i}
\mathcal{L}_{kl}p_{k}+\frac{N_{0}W}{M}}\right)\right],~~~k,l\in \mathbb{C}_{j}, l\neq i,
\end{eqnarray}
with the expectation is computed with respect to
$\textbf{P}^{(j)}_{-i}$ and $\lbrace \mathcal{L}_{kl}
\rbrace_{_{k,l\in \mathbb{C}_{j}}}$ excluding $l=i$\footnote{Note that the power of the users are random variables, since they are a deterministic function of their corresponding  direct channel gains, which are random variables.}. It is worth mentioning that the power
$p_{i}$ in (\ref{term2}) prevents the $i$th user from selfishly
maximizing its average rate given in (\ref{term1}). Using the fact
that all users follow the same power allocation policy, and since
the channel gains $\mathcal{L}_{kl}$ are random variables with the
same distributions, $\bar{R}_{l}(p_{i})$ becomes independent of $l$.
Thus, by dropping the index $l$ from $\bar{R}_{l}(p_{i})$, the
utility function of link $i$ can be simplified as
\begin{equation}\label{DPA1}
u_{i}(p_{i},h_{ii})=\bar{R}_{i}(p_{i},h_{ii})+(n-1)\bar{R}(p_{i}).
\end{equation}
Noting that $p_{i}$ depends only on the channel gain $h_{ii}$, in
the sequel we use $p_{i}=g(h_{ii})$.

\begin{lem} \label{lemma001} Let us assume $\mathbb{E}[p_{k}] \triangleq q_{n}$, $0<\alpha \leq 1$ 
is fixed and the interference is strong enough ($\mathbb{E} [I_i]= \omega (1)$). Then \textit{with 
probability one} (w. p. 1), we have
\begin{equation}
 I_{i}~\sim~ (n-1) \hat{\alpha} q_{n},
\end{equation}
as $K \rightarrow \infty $ (or equivalently, $n \to \infty$), where $\hat{\alpha} \triangleq \alpha \varpi$. 
More precisely, substituting $I_i$ by $(n-1) \hat{\alpha} q_{n}$ does not change the asymptotic average 
sum-rate of the network.
\end{lem}
\begin{proof}
See Appendix \ref{append01}.
\end{proof}

\begin{lem} \label{lemma002}
For large values of $n$, the links with a direct channel
gain above $h_{Th}=c \log n$, where $c>1$ is a constant, 
have negligible contribution in the network average sum-rate. 
\end{lem}
\begin{proof}
See Appendix \ref{append02}.
\end{proof}
%Since $\log (1+x) \leq x$, for $x > 0$, the above argument is valid for  $\mathbb{E}[\Xi_{i}(\hat{p}_{i},h_{ii})]$. 

From Lemma \ref{lemma002} and for large values 
of $n$, we can limit our attention to a subset of links for which the direct channel gain $h_{ii}$ is 
less than $c\log n$, $c>1$.

\begin{thm}\label{th1} Assuming the strong interference scenario and sufficiently large $K$, 
the optimum power allocation policy for (\ref{DPA}) is $\hat{p}_{i}=g(h_{ii})=U(h_{ii}-\tau_{n})$, where
$\tau_{n}>0$ is a threshold level which is a function of $n$, and $U(.)$ is the unit 
step function. Also, the maximum network average sum-rate in (\ref{eqn: 05}) is achieved at 
$M=1$ and is given by 
\begin{equation} \label{er}
\bar{R}_{ave} \sim  \frac{W}{\hat{\alpha}} \log K .
\end{equation}
\end{thm}
\begin{proof}
The steps of the proof are as follows: First, we derive an upper
bound on the utility function given in (\ref{DPA1}). Then, we prove
that the optimum power allocation strategy that maximizes this upper
bound is $\hat{p}_{i}=g(h_{ii})=U(h_{ii}-\tau_{n})$. Based on this
power allocation policy, in Lemma \ref{lemma004}, we derive the
optimum threshold level $\tau_{n}$. We then show that using this optimum threshold value, the maximum value of the
utility function in (\ref{DPA1}) becomes asymptotically
the same as the maximum value of the upper bound obtained in the
first step. Finally, the proof of the theorem is completed by showing that
the maximum network average sum-rate is achieved at $M=1$.

\underline{\textbf{Step 1:} \textit{Upper Bound on the Utility Function}}

Let us assume $\mathbb{E}\left[ p_{k}\right]=q_{n} $. Using the results of Lemma
\ref{lemma001}, $\bar{R}_{i}(p_{i},h_{ii})$ in (\ref{DPA1}) can be
expressed as
\begin{eqnarray}
\bar{R}_{i}(p_{i},h_{ii}) & \approx & \dfrac{W}{M} \mathbb{E} \left[ \log \left(1+\dfrac{h_{ii}p_{i}}{(n-1)\hat{\alpha}q_{n}+\frac{N_{0}W}{M}}\right) \right]\\
\label{term112} &\stackrel{(a)}{=}&\dfrac{W}{M} \log\left( 1+\dfrac{h_{ii} p_{i}}{\lambda}\right),
\end{eqnarray}
as $K \rightarrow \infty $, where
\begin{equation}\label{landa}
\lambda \triangleq (n-1) \hat{\alpha}q_{n}+\frac{N_{0}W}{M}.
\end{equation}
In the above equations, $(a)$ follows from the fact that $h_{ii}$ is
a known parameter for user $i$ and $p_{i} = g(h_{ii})$ is the optimization
parameter.  With a similar
argument, (\ref{term2}) can be simplified as
\begin{eqnarray}
\bar{R} (p_{i}) & \approx & \dfrac{W}{M}\mathbb{E} \left[\log\left( 1+\dfrac{h_{ll}p_{l}}{\mathcal{L}_{il}p_{i}+(n-2)\hat{\alpha}q_{n}+\frac{N_{0}W}{M}}\right) \right],~~i \neq l\\
\nonumber &\stackrel{(a)}{=}& \alpha \dfrac{W}{M}\mathbb{E} \left[ \log\left( 1+\dfrac{h_{ll}p_{l}}{\beta_{il} h_{il}p_{i}+(n-2)\hat{\alpha}q_{n}+\frac{N_{0}W}{M}}\right)  \right]+\\
\label{term221} && (1-\alpha) \dfrac{W}{M}\mathbb{E} \left[ \log\left( 1+\dfrac{h_{ll}p_{l}}{(n-2)\hat{\alpha}q_{n}+\frac{N_{0}W}{M}}\right)\right]\\
\label{term222}&=&\dfrac{\alpha W}{M} \mathbb{E} \left[\log\left( 1+\dfrac{h_{ll}p_{l}}{\beta_{il} h_{il}p_{i}+\lambda'} \right) \right]+(1-\alpha)\dfrac{W}{M}\mathbb{E}  \left[\log\left( 1+\dfrac{h_{ll}p_{l}}{\lambda'} \right) \right],
\end{eqnarray}
as $K \rightarrow \infty $, where the expectation is computed with respect to $h_{ll}$, $h_{il}$, 
$p_{l}$ and $\beta_{il}$, and $\lambda' \triangleq (n-2) \hat{\alpha} q_n +\frac{N_{0}W}{M}$. Also, 
$(a)$ comes from the shadowing model described in (\ref{eqn: 01}). Using
(\ref{term112}), (\ref{term222}), and the inequality $\log (1+ x)
\leq x$, the utility function in (\ref{DPA1}) is upper bounded as\footnote{Note that the factor $(n-1)$ in (\ref{DPA1}) is replaced by $n$ in (\ref{ff}), which does not affect the validity of the equation.}
\begin{equation}\label{ff}
u_{i}(p_{i},h_{ii}) \leq \dfrac{W}{M}\dfrac{h_{ii}}{\lambda}p_{i} +n\dfrac{\alpha W}{M} \mathbb{E}  \left[\dfrac{h_{ll}p_{l}}{\beta_{il} h_{il}p_{i}+\lambda'} \right]+n(1-\alpha)\dfrac{W }{M \lambda'}\mathbb{E} \left[ h_{ll}p_{l} \right].
\end{equation}
Noting that $h_{ll}$ is independent of $h_{il}$, $i \neq l$, we have
\begin{eqnarray}\label{term4f}
\mathbb{E}\left[\dfrac{h_{ll}p_{l}}{\beta_{il} h_{il}p_{i}+\lambda'} \Big \vert \beta_{il} \right] & = & \mu  \int_{0}^{\infty}\dfrac{e^{-y}}{y\beta_{il}p_{i}+\lambda'}dy \\
\label{hg1}& = & - \dfrac{\mu}{\beta_{il} p_{i}} e^{\frac{\lambda'}{\beta_{il} p_{i}}} \mathrm{Ei} \left(-\frac{\lambda'}{\beta_{il} p_{i}}\right),
\end{eqnarray}
where
\begin{equation}\label{mu1}
\mu \triangleq \mathbb{E} \left[ h_{ll}p_{l} \right],
\end{equation}
 and
$\mathrm{Ei}(x) \triangleq -\int_{-x}^{\infty}\frac{e^{-t}}{t}dt$, $x<0$
is the \textit{exponential-integral function} \cite{Gradshteynbook94}.
Thus, the right hand side of (\ref{ff}) is simplified as
\begin{equation}\label{eq: a1}
u_{i}(p_{i},h_{ii}) \leq \dfrac{W}{M}\dfrac{h_{ii}}{\lambda}p_{i} - n\dfrac{\alpha \mu W}{M}\mathbb{E}\left[\dfrac{1}{\beta_{il} p_{i}} e^{\frac{\lambda'}{\beta_{il} p_{i}}} \mathrm{Ei} \left(-\frac{\lambda'}{\beta_{il} p_{i}}\right)\right]+n(1-\alpha) \dfrac{W}{M}\dfrac{\mu}{\lambda'},
\end{equation}
where the expectation is computed with respect to $\beta_{il}$.
An asymptotic expansion of $\mathrm{Ei}(x)$ can be obtained as \cite[p. 951]{Gradshteynbook94}
\begin{equation}\label{gamma2}
\mathrm{Ei}(x)=\dfrac{e^{x}}{x}\left[ \sum_{k=0}^{L-1} \dfrac{k!}{x^k}+O(\vert x  \vert^{-L})\right];~ ~L=1,2,...,
\end{equation}
as $ x \rightarrow - \infty $. Setting $L=4$, we can rewrite (\ref{eq: a1}) as
\begin{eqnarray}
\nonumber u_{i}(p_{i},h_{ii}) &\leq& \dfrac{W}{M}\dfrac{h_{ii}}{\lambda}p_{i} + n\dfrac{\alpha W \mu}{M\lambda'} \mathbb{E}\left[ \left( 1-\dfrac{\beta_{il} p_{i}}{\lambda'}+2\left( \dfrac{\beta_{il} p_{i}}{\lambda'}\right)^{2}-6\left( \dfrac{\beta_{il} p_{i}}{\lambda'}\right)^{3}\right)\right]+\\
&& n\dfrac{\alpha W \mu}{M\lambda'} \mathbb{E}\left[O\left(\Big \vert \dfrac{\beta_{il} p_{i}}{\lambda'}  \Big \vert^{4}\right)\right]+ n(1-\alpha)\dfrac{W \mu}{M\lambda'}\\
\label{asy01}&\stackrel{(a)}{\approx}& \dfrac{W}{M}\dfrac{h_{ii}}{\lambda}p_{i} + n\dfrac{\alpha W \mu}{M\lambda'} \left( 1-\dfrac{\varpi p_{i}}{\lambda'}+2\kappa\left( \dfrac{ p_{i}}{\lambda'}\right)^{2}-6\eta\left( \dfrac{ p_{i}}{\lambda'}\right)^{3}\right)+n(1-\alpha)\dfrac{W \mu}{M\lambda'}\\
&\triangleq& \Xi_{i}(p_{i},h_{ii}),
\end{eqnarray}
as $\lambda' \rightarrow \infty$, where $ \kappa \triangleq \mathbb{E} \big[\beta_{ki}^{2}\big]$ and $\eta \triangleq \mathbb{E} \big[\beta_{ki}^{3}\big]$, and $(a)$ follows from the fact
that for large values of $\lambda'$, the term $\mathbb{E}\left[O\left(\Big
\vert \dfrac{\beta_{il} p_{i}}{\lambda'}  \Big
\vert^{4}\right)\right]$ can be ignored. 

%As we will see later, the condition $\lambda \rightarrow \infty$ is satisfied for the optimum power allocation policy.

\underline{\textbf{Step 2:} \textit{Optimum Power Allocation Policy for $\Xi_{i}(p_{i},h_{ii})$}}

Using the fact that $p_{i} \in [0,1]$, the second-order derivative
of (\ref{asy01}) in terms of $p_{i}$, $\dfrac{\partial^{2}
\Xi_{i}(p_{i},h_{ii})}{\partial p^{2}_{i}}=n\dfrac{\alpha W
\mu}{M\lambda'}\left( \dfrac{4\kappa}{\lambda'^2}-\dfrac{36
\eta}{\lambda'^{3}}p_{i} \right)$, is positive\footnote{It is 
observed from (\ref{gamma2}) and (\ref{asy01}) that for any 
value of $L >4$, the second-order derivative
of (\ref{asy01}) in terms of $p_{i}$ is positive too.} as $\lambda'
\rightarrow \infty$. Thus, (\ref{asy01}) is a \textit{convex}
function of $p_{i}$. It is known that a convex function attains its
maximum at one of its extreme points\footnote{In the power domain
$\mathscr{P}=[0,1]$, the extreme points are $0$ and $1$.} of its
domain \cite{Bertsekas1999}. In other words, the optimum power that maximizes
(\ref{asy01}) is $\hat{p}_{i} \in \lbrace 0,1\rbrace$. To show that
this optimum power is in the form of a unit step function, it is
sufficient to prove that $p_i=g(h_{ii})$ is a monotonically increasing
function of $h_{ii}$. 

Suppose that the optimum power that maximizes
$\Xi_{i}(p_{i},h_{ii})$ is $p_{i}=1$. Also, let us define
$h_{ii}^{'} \triangleq h_{ii}+\delta$, where $\delta >0$. From
(\ref{asy01}), it is clear that $\Xi_{i}(p_{i},h_{ii})$ is a
monotonically increasing function of $h_{ii}$, i.e.,
\begin{equation}\label{eqn: monotical01}
\Xi_{i}(p_{i}=1,h^{'}_{ii})>\Xi_{i}(p_{i}=1,h_{ii}).
\end{equation}
 On the other hand, since
the optimum power is $p_{i}=1$, we conclude that
\begin{equation}\label{eqn: monotical02}
\Xi_{i}(p_{i}=1,h_{ii})>\Xi_{i}(p_{i}=0,h_{ii}).
\end{equation}
Using the fact that $\Xi_{i}(p_{i}=0,h_{ii})=\Xi_{i}(p_{i}=0,h^{'}_{ii})$, we arrive at the following inequality
\begin{equation}\label{eqn: monotical03}
\Xi_{i}(p_{i}=1,h^{'}_{ii})>\Xi_{i}(p_{i}=0,h^{'}_{ii}).
\end{equation}

From (\ref{eqn: monotical01})-(\ref{eqn: monotical03}), it is
concluded that $g(h_{ii})$ is a monotonically increasing function of
$h_{ii}$. Consequently, the optimum power allocation strategy that
maximizes $\Xi_{i}(p_{i},h_{ii})$ is a unit step function, i.e.,
\begin{eqnarray} \label{eq: 114}
    \hat{p}_{i}=
\left\{\begin{array}{ll}
1   ,    & \textrm{if}~~ h_{ii}> \tau_{n} \\
0 ,  &  \textrm{Otherwise},
\end{array} \right.
\end{eqnarray}
where $\tau_{n}$ is a threshold level to be determined. We call this the \textit{threshold-based 
on-off power allocation strategy}. It is
observed that the optimum power $\hat{p}_{i}$ is a Bernoulli random
variable with parameter $q_{n}$, i.e.,
\begin{eqnarray}\label{eq: 14}
    f(\hat{p}_{i})=
\left\{\begin{array}{ll}
q_{n}   ,    &  \hat{p}_{i}=1,  \\
1-q_{n} ,  &  \hat{p}_{i}=0,
\end{array} \right.
\end{eqnarray}
where $f(.)$ is the probability mass function (pmf) of
$\hat{p}_{i}$. We conclude from (\ref{eq: 114}) and (\ref{eq: 14})
that the probability of link activation in each cluster is $q_{n}
\triangleq \mathbb{P} \left\lbrace h_{ii} > \tau_{n}\right\rbrace=e^{-\tau_{n}}$
which is a function of $n$.

\underline{\textbf{Step 3:} \textit{Optimum Threshold Level $\tau_{n}$}}

From Step 1, it is observed that for every value of $p_{i}$ we have
\begin{equation}
 u_{i}(p_{i},h_{ii}) \leq \Xi_{i}(p_{i},h_{ii}).
\end{equation}
The above inequality is also valid for the optimum power $\hat{p}_{i}$
obtained in Step 2. Thus, using the fact that for $X \leq Y$,
$\mathbb{E}[X] \leq \mathbb{E}[Y]$, we conclude
\begin{equation}
\mathbb{E}[u_{i}(\hat{p}_{i},h_{ii})] \leq  \mathbb{E}[\Xi_{i}(\hat{p}_{i},h_{ii})],
\end{equation}
where the expectations are computed with respect to $h_{ii}$. In the
following lemmas, we first derive the optimum threshold level $\tau_{n}$
that maximizes $\mathbb{E}[\Xi_{i}(\hat{p}_{i},h_{ii})]$, and then prove that this
quantity is asymptotically the same as the optimum threshold level
maximizing\footnote{In fact, since the threshold $\tau_n$ is fixed
and does not depend on a specific realization of $h_{ii}$, finding
the optimum value of $\tau_n$ requires averaging the utility
function over all realizations of $h_{ii}$.}
$\mathbb{E}[u_{i}(\hat{p}_{i},h_{ii})]$, assuming an on-off power scheme.
We also show that the maximum value of
$\mathbb{E}[u_{i}(\hat{p}_{i},h_{ii})]$ (assuming an on-off power scheme)
is the same as the optimum value of
$\mathbb{E}[\Xi_{i}(\hat{p}_{i},h_{ii})]$, proving the desired
result.

\begin{lem} \label{lemma003}
For large values of $n$ and given $0<\alpha \leq 1$, the optimum
threshold level that maximizes
$\mathbb{E}[\Xi_{i}(\hat{p}_{i},h_{ii})]$ is computed as
\begin{equation}
 \hat{\tau}_{n} \sim \log n.
\end{equation}
Also, the maximum value of $\mathbb{E}[\Xi_{i}(\hat{p}_{i},h_{ii})]$
scales as $\dfrac{W}{M \hat{\alpha}}\log n $.
\end{lem}
\begin{proof}
See Appendix \ref{append03}.
\end{proof}

\begin{lem} \label{lemma004}
For large values of $n$ and given $0<\alpha \leq 1$,

i) The optimum threshold level that maximizes
$\mathbb{E}[u_{i}(\hat{p}_{i},h_{ii})]$ is computed as
\begin{equation}\label{hth42}
\hat{\tau}_{n}=\log n-2\log \log n + O(1),
\end{equation}

ii) The probability of link activation in each cluster is given by
\begin{equation}\label{eqn: 15}
q_{n}  = \delta \dfrac{ \log ^{2} n }{n},
\end{equation}
where $\delta > 0$ is a constant,

iii) The maximum value of $\mathbb{E}[u_{i}(\hat{p}_{i},h_{ii})]$ scales as $\dfrac{W}{M\hat{\alpha}}\log n $.
\end{lem}
\begin{proof}
See Appendix \ref{append04}.
\end{proof}

\underline{\textbf{Step 4:} \textit{Optimum Power Allocation Strategy that Maximize $u_{i}(p_{i},h_{ii})$}}

In order to prove that the utility function in (\ref{DPA1}) is asymptotically the
same as the upper bound $\Xi_{i}(p_{i},h_{ii})$ obtained in (\ref{asy01}), it is 
sufficient to show that the low SINR conditions in (\ref{term112}) and (\ref{term222}) 
are satisfied. Using (\ref{term112}), (\ref{landa}) and (\ref{eqn: 15}), the SINR is
equal to $\dfrac{h_{ii} p_{i}}{\lambda}$, where
\begin{equation}
\lambda \approx \hat{\alpha}\delta \log ^{2} n+\frac{N_{0}W}{M}.
\end{equation}
It is observed that $\lambda$ goes to infinity as $n \rightarrow
\infty$. On the other hand, since we are limiting our attention to
links with $h_{ii} < h_{Th}=c\log n$, we have
\begin{equation}
\frac{h_{ii}p_{i}}{\lambda}=O\left(\dfrac{1}{\log n}\right),
\end{equation}
when $n \rightarrow \infty$. Thus, for large values of $n$, the low
SINR condition, $\frac{h_{ii}p_{i}}{\lambda} \ll 1$, is satisfied.
With a similar argument, the low SINR condition for (\ref{term222})
is satisfied. Hence, we can use the approximation $\log(1+x) \approx
x$, for $x \ll 1$, to simplify (\ref{term112}) and
(\ref{term222}) as follows:
\begin{equation}
\bar{R}_{i}(p_{i},h_{ii}) \approx \dfrac{W}{M}\dfrac{h_{ii}}{\lambda}p_{i},
\end{equation}
\begin{equation}
\bar{R}(p_{i}) \approx \dfrac{\alpha W}{M} \mathbb{E}  \left[\dfrac{h_{ll}p_{l}}{\beta_{il} h_{il}p_{i}+\lambda'} \right]+(1-\alpha)\dfrac{W }{M \lambda'}\mathbb{E} \left[ h_{ll}p_{l} \right].
\end{equation}
Consequently, the utility function $u_{i}(p_{i},h_{ii})$ is the same
as the upper bound $\Xi_{i}(p_{i},h_{ii})$ obtained in
(\ref{asy01}), when $n \rightarrow \infty$. Thus, the optimum power
allocation strategy for (\ref{DPA}) is the same as the optimum power
allocation policy that maximizes $\Xi_{i}(p_{i},h_{ii})$.

\underline{\textbf{Step 5:} \textit{Maximum Average Network Sum-rate}}

Using (\ref{utility1}), the average utility function of each user
$i$, $\mathbb{E} \left[u_{i}(\hat{p}_{i},h_{ii}) \right], ~~i \in
\mathbb{C}_{j}$, is the same as the average sum-rate of the links in
cluster $\mathbb{C}_{j}$ represented by
\begin{equation}\label{fds}
\bar{R}_{ave}^{(j)}  \triangleq \sum_{i \in \mathbb{C}_{j}}
\mathbb{E}\left[R_{i}(\hat{\textbf{\textrm{P}}}^{(j)},\boldsymbol{\mathcal{L}}_{i}^{(j)})\right],~~j=1,...,M.
\end{equation}
where $\hat{\textbf{\textrm{P}}}^{(j)}$ is the on-off powers vector of the links in cluster $\mathbb{C}_{j}$. 
In this case, the network average sum-rate defined in (\ref{eqn: 05}) can be written as
\begin{eqnarray}
\label{sumrate1} \bar{R}_{ave} & = & \sum_{j=1}^{M} \bar{R}_{ave}^{(j)}\\
\label{fun01} &\stackrel{(a)}{\approx}& \dfrac{W \hat{\tau}_n}{\hat{\alpha} } ,
\end{eqnarray}
where $(a)$ follows from (\ref{eqn:q102}) of Appendix \ref{append04}. Using (\ref{hth42}),  and noting that $n=\dfrac{K}{M}$, we have
\begin{equation}\label{eqn:sumrate1}
\bar{R}_{ave} \sim \frac{W}{\hat{\alpha}} \log \frac{K}{M}. 
\end{equation}

\underline{\textbf{Step 6:} \textit{Optimum Spectrum Allocation}}

According to (\ref{fun01}), the network average sum-rate is a monotonically
increasing function of $\hat{\tau}_n$. Rewriting equation (\ref{eqn:d9}) of Appendix \ref{append04}, 
which gives the optimum threshold value for the on-off scheme:
\begin{eqnarray}
-e^{-\hat{\tau_n}} \log \left( 1+\frac{\hat{\tau}_n e^{\hat{\tau}_n}}{n \hat{\alpha}}\right) + \frac{1+\hat{\tau}_n}{n \hat{\alpha} + \hat{\tau}_n e^{\hat{\tau}_n}} =0,
\end{eqnarray}
it can be shown that\footnote{In deriving (\ref{eqn:d9_2}), we have used the fact that $\frac{\hat{\tau}_n e^{\hat{\tau}_n}}{n \hat{\alpha}} \ll 1$, which is feasible based on the solution given in (\ref{hth42}).}
\begin{eqnarray} \label{eqn:d9_2}
 \hat{\tau}_n^2 e^{\hat{\tau_n}} \approx n \hat{\alpha},
\end{eqnarray}
which implies that $\hat{\tau}_n$ is an increasing function of $n$. Therefore, the average sum-rate 
of the network is an increasing function of $n$ and consequently, noting that $n=\frac{K}{M}$, is a decreasing 
function of $M$. Hence, the maximum average sum-rate  of the network for the strong interference 
scenario and $0 < \alpha <1$ is obtained at $M=1$ and this completes the proof of the theorem.
\end{proof}

Motivated by Theorem \ref{th1}, we describe the proposed
threshold-based on-off power allocation strategy for single-hop
wireless networks. Based on this scheme, all users perform the
following steps during each block:
\begin{itemize}
\item [1-] Based on the direct channel gain, the transmission policy is
\begin{eqnarray} \nonumber
    \hat{p}_{i}=
\left\{\begin{array}{ll}
1   ,    & \textrm{if}~~ h_{ii}> \tau_{n} \\
0 ,  &  \textrm{Otherwise}.
\end{array} \right.
\end{eqnarray}

\item [2-] Knowing its corresponding direct channel gain, each active user $i$ transmits with full power and rate
\begin{eqnarray}
 R_i = \log \left( 1+ \frac{h_{ii}}{(n-1)\hat{\alpha} e^{-\tau_n} + \frac{N_0 W}{M}}\right).
\end{eqnarray}

\item [3-] Decoding is performed over sufficiently large number of blocks, yielding the average rate of $\frac{W}{\hat{\alpha}K} \log K$ for each user, and the average sum-rate of $\frac{W}{\hat{\alpha}} \log K$ in the network.
\end{itemize}

\textbf{Remark 1-} Theorem \ref{th1} states that the average sum-rate of the network for fixed $M$ 
depends on the value of $\hat{\alpha} = \alpha \varpi$ and scales as
$\frac{W}{\hat{\alpha}} \log \frac{K}{M}$. Also, for values of $M$
such that $\log M = o(\log K )$, the network average sum-rate scales 
as $\frac{W}{\hat{\alpha}} \log K$.

\textbf{Remark 2-} Let $m_{j}$ denote the number of active links in $\mathbb{C}_{j}$. Lemma 4 states that the optimum selection of the threshold value yields  $\mathbb{E}[m_{j}]=nq_{n} = \Theta \left (\log ^{2} n
\right)$. More precisely, it can be shown that the optimum number of active users scales as $\Theta \left (\log ^{2} n
\right)$, with probability one.

\subsection{Moderate and Weak Interference Scenarios ($\mathbb{E}[I_i]=O(1)$)}

\begin{thm} \label{thm2}
Let us assume $K$ is large and $M$ is fixed. Then,

i) For the moderate interference (i.e., $\mathbb{E}[I_i]=\Theta(1)$), the network average sum-rate is bounded by $\bar{R}_{ave} \leq \Theta(\log n)$.

ii) For the weak interference (i.e., $\mathbb{E}[I_i]=o(1)$), the network average sum-rate is bounded by $\bar{R}_{ave} \leq o(\log n)$.
\end{thm}
\begin{proof}
i) From (\ref{eqn: 05}), we have
\begin{eqnarray}
\bar{R}_{ave} & = & \sum_{j=1}^{M}\sum_{l \in \mathbb{C}_{j}}\mathbb{E}\left[ \dfrac{W}{M} \log \left( 1+\dfrac{h_{ll}\hat{p}_{l}}{I_{l}+\frac{N_{0}W}{M}}\right) \right]\\
&\stackrel{(a)}{\leq}& \sum_{j=1}^{M}\sum_{l \in \mathbb{C}_{j}}\dfrac{W}{M}\mathbb{E}\left[  \log \left( 1+\dfrac{\hat{p}_{l} c \log n}{I_{l}+\frac{N_{0}W}{M}}\right) \right]\\
&\leq& \sum_{j=1}^{M}\sum_{l \in \mathbb{C}_{j}}\dfrac{W}{M}\mathbb{E}\left[  \log \left( 1+\dfrac{\hat{p}_{l} c \log n}{\frac{N_{0}W}{M}}\right) \right]\\
&\stackrel{(b)}{\leq}& \sum_{j=1}^{M}\sum_{l \in \mathbb{C}_{j}}\dfrac{W}{M} \log \left( 1+\dfrac{cq_{n} \log n}{\frac{N_{0}W}{M}}\right)\\
\label{eqn:inter01}&\stackrel{(c)}{\leq}& \dfrac{cM}{N_{0}} n q_{n}\log n,
\end{eqnarray}
where $(a)$ follows from Lemma 2, which implies that the realizations in which $h_{ll} > c \log n$ 
for some $c>1$ has negligible contribution in the network average sum-rate, $(b)$ results from the 
\textit{Jensen's inequality}, $\mathbb{E}\left[\log x
\right]\leq\log (\mathbb{E}\left[x \right])$, $x>0$. Also, $(c)$ follows from the 
fact that $\log (1+ x) \leq x$, $x>0$. Since for the moderate interference, 
$\mathbb{E}[I_{i}]=\hat{\alpha} nq_{n}=\Theta(1)$, and using the fact that 
$M$ is fixed, we come up with the following inequality
\begin{eqnarray}
\bar{R}_{ave} &\leq& \dfrac{cM}{\hat{\alpha}N_{0}} \Theta(1) \log n \\
&=& \Theta (\log n).
\end{eqnarray}

ii) For the weak interference scenario, where $\mathbb{E}[I_{i}]=\hat{\alpha} nq_{n}=o(1)$, and 
similar to the part (i), it is concluded from (\ref{eqn:inter01}) that
\begin{eqnarray}
\bar{R}_{ave} &\leq& \dfrac{cM}{\hat{\alpha}N_{0}} o(1) \log n \\
&=& o (\log n).
\end{eqnarray}
\end{proof}

\textbf{Remark 3-} It is concluded from Theorems \ref{th1} and \ref{thm2} that 
the maximum average sum-rate of the proposed network is scaled as $\Theta(\log K)$.

\subsection{$M$ Not Fixed (Scaling With $K$)}
So far, we assume that $M$ is fixed, i.e., it does not scale with $K$. In the following, we present 
some results for the case that $M$ scales with $K$\footnote{Obviously, we consider the values of $M$ which are in the interval $[1,K]$.}. It should be noted that the results for $M=o(K)$
is the same as the results in Theorem \ref{th1}.
 
\begin{thm} \label{corol004} In the network with the on-off power allocation strategy, 
if $M = \Theta(K) $ and $0<\alpha<1$, then the maximum network average sum-rate in 
(\ref{eqn: 05}) is less than that of $M=1$. Consequently, the maximum average sum-rate 
of the network for every value of $1 \leq M \leq K$ is achieved at $M=1$.
\end{thm}
\begin{proof}
See Appendix \ref{append05}.
\end{proof}

\textbf{Remark 4-} According to the shadow-fading model proposed in (\ref{eqn: 01}), it 
is seen that for $\alpha=0$, with probability one, $\mathcal{L}_{ki}=0,~k\neq i$. This 
implies that no interference exists in each cluster. In this case, the maximum average 
sum-rate of the network is clearly achieved by all users in the network transmitting at 
full power. It can be shown that for every value of $1 \leq M \leq K$, the maximum network 
average sum-rate for $\alpha = 0$ is achieved at $M=1$ (See Appendix \ref{append06} for the proof).

\textbf{Remark 5-} Noting that for $M=K$ only one user exists in
each cluster, all the users can communicate using an interference
free channel. It can be shown that for $M=K$ and every value of
$0 \leq \alpha \leq 1$, the network average sum-rate is asymptotically
obtained as 
\begin{equation}\label{eqn: networkthroughputM=K}
\bar{R}_{ave} \approx W(\log K-\log N_{0}W-\boldsymbol{\gamma}),
\end{equation}
where $\boldsymbol{\gamma}$ is Euler's constant (See Appendix \ref{append07} for the proof). 
Therefore, for every value of $0<\alpha<1$, it is observed that the average sum-rate of the network in 
(\ref{eqn: networkthroughputM=K}) is less than that of $M=1$ obtained in (\ref{er}).

\textbf{Remark 6-} Note that for $M=1$, in which the average number of active links scales 
as $\Theta(\log ^{2} K)$ (in the optimum on-off scheme), we have significant energy saving in the 
network as compared to the case of $M=K$, in which all the users transmit with full power.

\subsection{Numerical Results}
\begin{figure}[bhpt]
\centerline{\psfig{figure=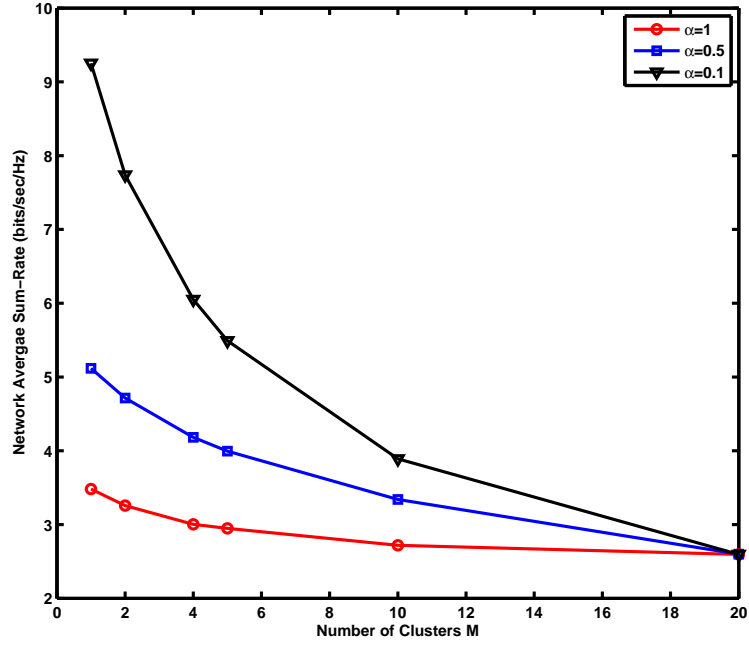,width=4.6in}}
\vspace{-7pt} \center{\hspace{16pt} \small{(a)}} \vspace{10pt}
\hspace{1pt} \centerline{\psfig{figure=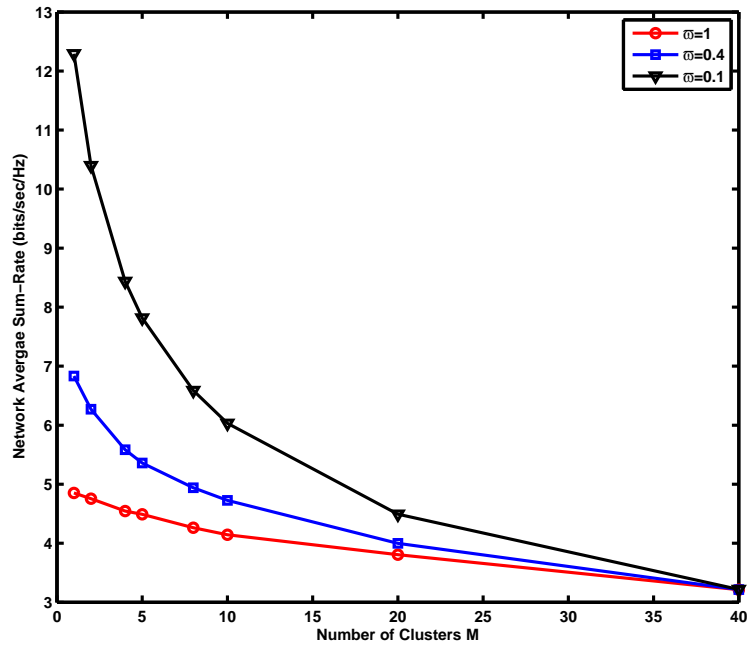,width=4.6in}}
\vspace{-35pt}
\center{\hspace{14pt} \small{(b)}} \\
\vspace{-7pt} \caption[a) $\lambda_{\ell}$ and b) Throughput of the
network vs. the number of links in each cluster for
$\Lambda_{1}(n)=\log n$ and $\Lambda_{2}(n)=e^{1.6\sqrt {\log n}}$ and
for $\hat{\alpha}=W=1$.] {Network average sum-rate vs. $M$ for \small{a) $K=20$, $\alpha=1,~0.5,~0.1$ and shadowing model with $\varpi=0.5$ and variance 1, and b) $K=40$, $\alpha=0.5$ and shadowing model with $\varpi=1,~0.4,~0.1$ and variance 1.}} 
\label{fig: sumrateM}
\end{figure}

So far, we have analyzed the average sum-rate of the network in terms of $M$ and 
$\hat{\alpha}$, in the asymptotic case of $K \to \infty$. For finite number of users, we 
have evaluated the network average sum-rate versus the number of clusters ($M$) 
through simulation. For this case, we assume that all the users in the network 
follow the threshold-based on-off power allocation policy, using the optimum 
threshold value. In addition, the shadowing effect is assumed to be \textit{lognormal} 
distributed with mean $\varpi \leq 1$ and variance 1. Fig. \ref{fig: sumrateM} shows 
the average sum-rate of the network versus $M$ for $K = 20$ and $K=40$, and 
different values of $\alpha$ and $\varpi$. It is observed from this figure that 
the average sum-rate of the network is a monotonically decreasing function of 
$M$ for every value of $(\alpha,\varpi)$, which implies that the maximum value 
of $\bar{R}_{ave}$ is achieved at $M = 1$. 

\begin{figure}[bhpt]
\centerline{\psfig{figure=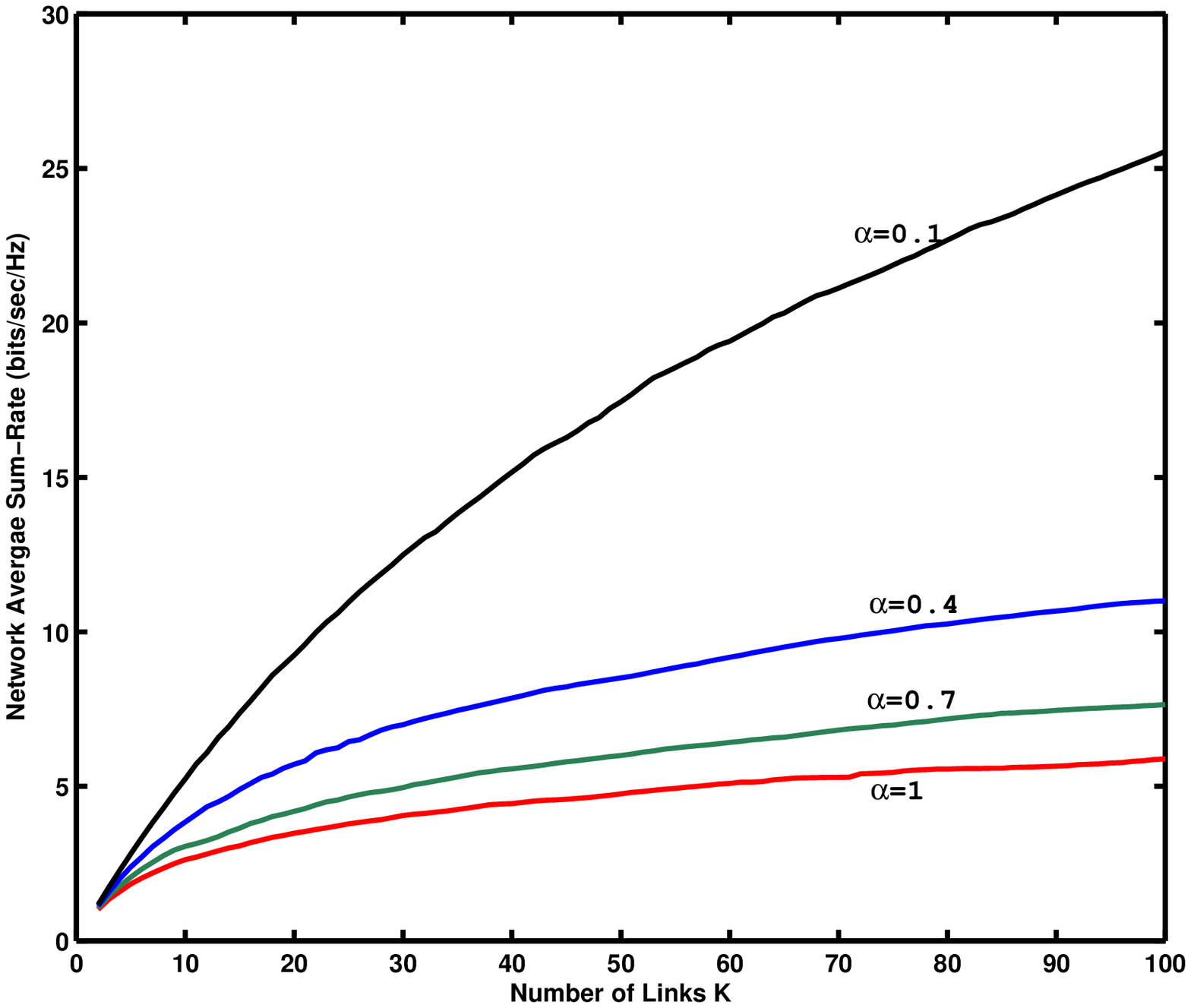,width=4.6in}}
\vspace{-7pt} \center{\hspace{16pt} \small{(a)}} \vspace{10pt}
\hspace{1pt} \centerline{\psfig{figure=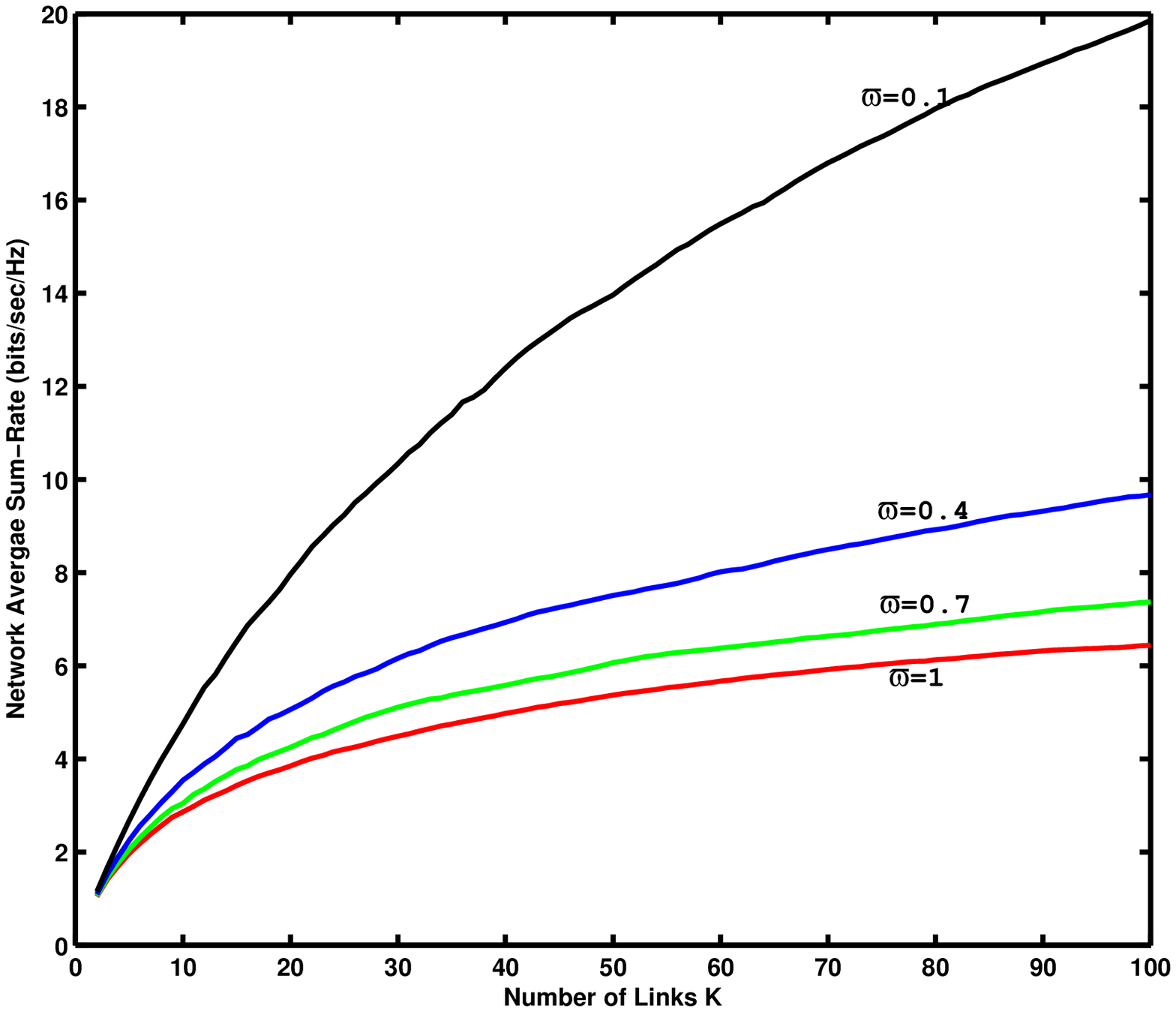,width=4.6in}}
\vspace{-35pt}
\center{\hspace{14pt} \small{(b)}} \\
\vspace{-7pt} \caption[a) $\lambda_{\ell}$ and b) Throughput of the
network vs. the number of links in each cluster for
$\Lambda_{1}(n)=\log n$ and $\Lambda_{2}(n)=e^{1.6\sqrt {\log n}}$ and
for $\hat{\alpha}=W=1$.] {Network average sum-rate vs. $K$ for $M=1$ and \small{a) shadowing model with $\varpi=0.5$ and variance 1, and $\alpha=1,~0.7,~0.4,~0.1$, and b) shadowing model with $\varpi=1,~0.7,~0.4,~0.1$ and variance 1, and $\alpha=0.5$.}} 
\label{fig: sumrate1}
\end{figure}

Based on the above arguments, we have plotted the average sum-rate of the network versus $K$ 
for $M=1$ and different values of $(\alpha,\varpi)$. It is observed from Fig. \ref{fig: sumrate1} 
that the network average sum-rate depends strongly on the values of $(\alpha,\varpi)$.

%\begin{figure}[t]
%\centerline{\psfig{figure=sumrate1.eps,width=4in}}
%\caption{Network average sum-rate vs. $K$ for $M=1$ and different values of $\alpha$.}
%\label{fig: sumrate1}
%\end{figure}

%\begin{figure}[t]
%%\centerline{\psfig{figure=sumrate2.eps,width=4in}}
%\caption{Network average sum-rate vs. $K$ for $M=1$ and different values of $\varpi$.}
%\label{fig: sumrate2}
%\end{figure}

\section{Network Guaranteed Sum-Rate}\label{guaranteed}
Recalling the definition of the network guaranteed sum-rate in (\ref{gau}), in this section 
we aim to find the maximum achievable guaranteed sum-rate of the network, as well as the 
optimum power allocation scheme and the optimum value of $M$. 

\begin{thm} \label{thmgau}
The guaranteed sum-rate of the underlying network in the asymptotic case of $K \to \infty$ is 
obtained by
\begin{eqnarray}
 \bar{R}_g \sim \frac{W}{\hat{\alpha}} \log K,
\end{eqnarray}
 which is achievable by the decentralized on-off power allocation scheme.
\end{thm}
\begin{proof}
In order to compute the guaranteed rate for link $l \in \mathbb{C}_j$, 
we first define the corresponding outage event as follows:
\begin{eqnarray}
\mathcal{O}_l^{(j)} &\equiv& \left\{ R_{l}(\textbf{\textrm{P}}^{(j)},\boldsymbol{\mathcal{L}}_{l}^{(j)}) <  R(h_{ll}) \right\} \\
&\equiv& \left \{ \log \left( 1+\frac{p_l h_{ll}}{I_l + \frac{N_0 W}{M}}\right) <  R(h_{ll}) \right\}.
\end{eqnarray}
In the following, we give an upper-bound and a lower-bound for $\bar{R}_g$ and show that these 
bounds converge to each other as $K \to \infty$ (or equivalently, $n \to \infty$). 

\textbf{Upper-bound:} An upper-bound on the guaranteed sum-rate can be given by lower-bounding the 
outage probability as follows:
\begin{eqnarray} 
 \mathbb{P} \left\{ \mathcal{O}_l^{(j)} \right\} &\geq& \mathbb{P} \left \{ \frac{p_l h_{ll}}{I_l + \frac{N_0 W}{M}} <  R(h_{ll}) \right\} \\
\label{gauranteed1}&=& \mathbb{P} \left \{ p_l h_{ll} - \frac{N_0 W}{M}  R(h_{ll})   <  I_l R(h_{ll}) \right\},
\end{eqnarray}
in which we have used the fact that $\log (1+x) \leq x$. Denoting $\nu = h_{ll}$, we can write
\begin{eqnarray} 
\mathbb{P} \left\{ \mathcal{O}_l^{(j)} \right\} &\stackrel{(a)}{\geq}& \mathbb{P} \left\{ e^{-I_l \xi (\nu) R (\nu) } \leq e^{\xi(\nu ) \left(\frac{N_0 W}{M}  R(\nu) -p_l \nu \right)} \right\} \\
\label{gau3}&\stackrel{(b)}{\geq}& 1- e^{- \xi(\nu ) \left(\frac{N_0 W}{M}  R(\nu) -p_l \nu \right)} \mathbb{E} \left[ e^{-I_l \xi (\nu) R (\nu) }\right],
\end{eqnarray}
for some positive $\xi (\nu)$. In the above equation, $(a)$ results from (\ref{gauranteed1}), 
noting that $\xi (\nu) >0$, and $(b)$ follows from Markov's inequality \cite[p. 77]{Rose2003}, and the expectation is taken with respect to $I_l$. The above equation implies 
that finding an upper-bound for $\mathbb{E} \left[ e^{-I_l \xi (\nu) R (\nu) }\right]$ is sufficient 
for the lower-bounding the outage probability. For this purpose, using (\ref{interf1}), we can write
\begin{eqnarray} 
\mathbb{E} \left[ e^{-I_l \xi (\nu) R (\nu) }\right] &=& \mathbb{E} \left[ e^{-\xi (\nu) R (\nu) \sum_{\substack{k  \in \mathbb{C}_{j} \\ k\neq l}} \mathcal{L}_{kl}p_{k}}\right] \\
&\stackrel{(a)}{=}& \prod_{\substack{k  \in \mathbb{C}_{j} \\ k\neq l}} \mathbb{E} \left[ e^{-\xi (\nu) R (\nu)  \mathcal{L}_{kl}p_{k}}\right] \\
&\stackrel{(b)}{=}& \prod_{\substack{k  \in \mathbb{C}_{j} \\ k\neq l}} \mathbb{E} \left[ e^{-\xi (\nu) R (\nu)  u_{kl} \beta_{kl} h_{kl}p_{k}}\right] \\
\label{gau4}&\stackrel{(c)}{=}& \left( \mathbb{E} \left[ e^{-\xi (\nu) R (\nu)  u_{kl} \beta_{kl} h_{kl}p_{k}}\right] \right)^{n-1},~k \neq l.
\end{eqnarray}
In the above equation, $(a)$ follows from the fact that $\{\mathcal{L}_{kl}\}_{k \in \mathbb{C}_j}$ with $k \neq l$,
and $\{p_k\}_{k \in \mathbb{C}_j}$ are mutually independent random variables, $(b)$ results from 
writing $\mathcal{L}_{kl}$ as $u_{kl} \beta_{kl} h_{kl}$ (from (\ref{eqn: 01})), in which $u_{kl}$ is 
an indicator variable which takes zero when $\mathcal{L}_{kl}=0$ and one, otherwise. $(c)$ follows 
from the symmetry which incurs that all the terms $\mathbb{E} \left[ e^{-\xi (\nu) R (\nu)  u_{kl} 
\beta_{kl} h_{kl}p_{k}}\right]$, $k \in \mathbb{C}_j$, are equal. Noting that $u_{kl}$, $\beta_{kl}$, 
$h_{kl}$, and $p_k$ are independent of each other, we have 
\begin{eqnarray} 
\mathbb{E} \left[ e^{-\xi (\nu) R (\nu)  u_{kl} \beta_{kl} h_{kl}p_{k}}\right] &=& \mathbb{E}_{\beta_{kl}} \left[\mathbb{E}_{h_{kl}} \left[\mathbb{E}_{u_{kl}} \left[ \mathbb{E}_{p_k} \left[ e^{-\xi (\nu) R (\nu)  u_{kl} \beta_{kl} h_{kl}p_{k}}\right]\right]\right]\right] \\
&\stackrel{(a)}{\leq}&  \mathbb{E}_{\beta_{kl}} \left[\mathbb{E}_{h_{kl}} \left[\mathbb{E}_{u_{kl}} \left[ (1-q_n) + q_n e^{-\xi (\nu) R (\nu)  u_{kl} \beta_{kl} h_{kl}}\right]\right]\right] \\
&\stackrel{(b)}{=}& \mathbb{E}_{\beta_{kl}} \left[\mathbb{E}_{h_{kl}} \left[   (1-q_n) + q_n \left( 1-\alpha + \alpha e^{-\xi (\nu) R (\nu)  \beta_{kl} h_{kl}} \right) \right] \right] \\
&\stackrel{(c)}{=}& \mathbb{E}_{\beta_{kl}} \left[ 1- \alpha q_n + \frac{\alpha q_n}{1+\beta_{kl} \xi (\nu) R (\nu)}\right] \\
&=& \mathbb{E}_{\beta_{kl}} \left[ 1- \frac{\alpha q_n \beta_{kl} \xi (\nu) R (\nu)}{1+\beta_{kl} \xi (\nu) R (\nu) }\right] \\
&\stackrel{(d)}{\leq}& 1- \frac{\alpha q_n \varpi \xi (\nu) R (\nu)}{1+ \beta_{\max} \xi (\nu) R (\nu)} \\
\label{gau5}&\stackrel{(e)}{\leq}& e^{-\frac{\hat{\alpha} q_n \xi (\nu) R (\nu)}{1+ \beta_{\max} \xi (\nu) R (\nu)}}.
\end{eqnarray}
In the above equation, $(a)$ follows from the fact that $e^{-\theta x} \leq (1-x) +x e^{-\theta}$, 
$\forall \theta \geq 0$ and $0 \leq x  \leq 1$, noting that $\mathbb{E} [p_k]=q_n$. $(b)$ results 
from the definition of $u_{kl}$, which is an indicator variable taking zero with probability 
$1-\alpha$ and one, with probability $\alpha$. $(c)$ follows from the fact that as $h_{kl}$ is 
exponentially-distributed, we have $\mathbb{E}_{h_{kl}} \left[ e^{-\xi (\nu) R (\nu)  \beta_{kl} 
h_{kl}} \right] = \frac{1}{1+\beta_{kl} \xi (\nu) R (\nu)}$. $(d)$ results from the facts 
that $\beta_{kl} \leq \beta_{\max}$ and $\mathbb{E} [\beta_{kl}]=\varpi$. Finally, $(e)$ follows 
from the fact that $1-x \leq e^{-x}$, $\forall x$ and noting that $\alpha \varpi = \hat{\alpha}$.

Combining (\ref{gau4})  and (\ref{gau5}) and substituting into (\ref{gau3}) yields
\begin{eqnarray}
\mathbb{P} \left\{ \mathcal{O}_l^{(j)} \right\} &\geq& 1- e^{- \xi(\nu ) \left(\frac{N_0 W}{M}  R(\nu) -p_l \nu \right)} e^{-\frac{(n-1) \hat{\alpha} q_n \xi (\nu) R (\nu)}{1+ \beta_{\max} \xi (\nu) R (\nu)}} \\
&=& 1-e^{- \xi(\nu) R (\nu) \left( \frac{(n-1) \hat{\alpha} q_n }{1+ \beta_{\max} \xi (\nu) R (\nu) } + \frac{N_0 W}{M}\right) \left(1- \frac{t(\nu)}{R(\nu)} \right)},
\end{eqnarray}
where $t(\nu) \triangleq \frac{p_l \nu}{\frac{(n-1) \hat{\alpha} q_n }{1+ \beta_{\max} \xi (\nu) R (\nu) } + \frac{N_0 W}{M}}$. 

Consider the cases of $\mathbb{E}\{I_l\}=\omega (1)$ (strong interference) or $\mathbb{E}\{I_l\}=\Theta (1)$ (moderate interference). Let us define $\gamma \triangleq \min \left(1, \frac{M (n-1) q_n \hat{\alpha}}{N_0 W} \right)$. Setting $\xi (\nu) \triangleq \frac{\frac{\gamma}{2}\frac{N_0 W}{M}}{\beta_{\max} R (\nu) \left( (n-1) \hat{\alpha} q_n-\frac{\gamma}{2}\frac{N_0 W}{M}\right)}$, we have $\frac{(n-1) \hat{\alpha} q_n }{1+ \beta_{\max} \xi (\nu) R (\nu) } + \frac{N_0 W}{M} = (n-1) \hat{\alpha} q_n + (1-\frac{\gamma}{2}) \frac{N_0 W}{M}$, and as a result,
\begin{eqnarray}
\mathbb{P} \left\{ \mathcal{O}_l^{(j)} \right\} &\geq& 1-e^{-\frac{\frac{\gamma}{2}\frac{N_0 W}{M} \left[(n-1) \hat{\alpha} q_n + (1-\frac{\gamma}{2}) \frac{N_0 W}{M}\right]}{\beta_{\max}  \left[ (n-1) \hat{\alpha} q_n-\frac{\gamma}{2}\frac{N_0 W}{M}\right]} \left(1- \frac{t(\nu)}{R(\nu)} \right)} \\
&\geq& 1-e^{-\frac{\gamma N_0 W}{2M \beta_{\max}}\left(1- \frac{t(\nu)}{R(\nu)} \right)}.
\end{eqnarray}
Since $\frac{\gamma N_0 W}{2M \beta_{\max}} = \Theta (1)$, it follows that the necessary condition to have $\mathbb{P} \left\{ \mathcal{O}_l^{(j)} \right\} \to 0$ is having $R (\nu) \lesssim t(\nu) = \frac{p_l \nu}{(n-1) \hat{\alpha} q_n + (1-\frac{\gamma}{2}) \frac{N_0 W}{M}}$. In other words,
\begin{eqnarray}
 R^* (\nu) \lesssim  \frac{p_l \nu}{(n-1) \hat{\alpha} q_n + (1-\frac{\gamma}{2}) \frac{N_0 W}{M}},
\end{eqnarray}
which implies that $\bar{R}_g$ defined in (\ref{gau}) is upper bounded by
\begin{eqnarray} 
\bar{R}_g &\lesssim& nW \mathbb{E}_{\nu} \left[ \frac{p_l \nu}{(n-1) \hat{\alpha} q_n + (1-\frac{\gamma}{2}) \frac{N_0 W}{M}}\right] \\
\label{gau6}&=& \frac{n W\mathbb{E}_{\nu} \left[ p_l \nu \right]}{(n-1) \hat{\alpha} q_n + (1-\frac{\gamma}{2}) \frac{N_0 W}{M}}.
\end{eqnarray}
Now, defining $\Psi_{n} \triangleq \log n + 2 \log \log n$, we have
\begin{eqnarray} 
\mathbb{E}_{\nu} \left[ p_l \nu \right] &\leq& \mathbb{E} \left[ p_l \nu | \nu \leq \Psi_{n} \right] \mathbb{P} \{\nu \leq \Psi_{n}\} + \mathbb{E} \left[ p_l \nu | \nu > \Psi_{n} \right] \mathbb{P} \{\nu > \Psi_{n}\} \\
&\stackrel{(a)}{\leq}& q_n \Psi_{n} + \mathbb{E} \left[ \nu | \nu > \Psi_{n} \right] \mathbb{P} \{\nu > \Psi_{n} \} \\
\label{gau06}&\stackrel{(b)}{=}& q_n \Psi_{n} + (\Psi_{n} +1) e^{-\Psi_{n}} \\
\label{gau7}&\stackrel{(c)}{\sim}& q_n \log n.
\end{eqnarray}
In the above equation, $(a)$ comes from the facts that $$\mathbb{E} \left[ p_l \nu | \nu \leq \Psi_{n} \right] \mathbb{P} \{\nu \leq \Psi_{n}\} \leq \Psi_{n} \mathbb{E} \left[ p_l | \nu  \leq \Psi_{n} \right] \mathbb{P} \{\nu \leq \Psi_{n}\} \leq \Psi_{n} \mathbb{E} [p_l] = \Psi_{n}  q_n,$$ and $0 \leq p_{l} \leq 1$. $(b)$ results from the fact that $\nu$ is exponentially-distributed. $(c)$ follows from the facts that i)~as we are considering the strong and moderate interference scenarios, it yields that $(n-1) \hat{\alpha} q_n = \Omega (1)$, or equivalently, $q_n = \Omega (\frac{1}{n})$, and ii)~the term $(\Psi_{n} +1) e^{-\Psi_{n}}$ scales as $\frac{1}{n \log n}$ (due to the definition of $\Psi_{n}$) which is negligible with respect to the first term $q_n \Psi_{n}$. Combining (\ref{gau6}) and (\ref{gau7}) yields
\begin{eqnarray} 
\bar{R}_g &\lesssim& \frac{W n q_n \log n}{(n-1) \hat{\alpha} q_n + (1-\frac{\gamma}{2}) \frac{N_0 W}{M}}\\
&\lesssim& \frac{W}{\hat{\alpha}} \log n \\
\label{gau8}&\lesssim& \frac{W}{\hat{\alpha}} \log K.
\end{eqnarray}
In the case of weak interference, we have 
\begin{eqnarray}
\bar{R}_g &\leq& n W \frac{\mathbb{E} [p_l \nu]}{\frac{N_0 W}{M}} \\
&=& \frac{M n}{N_0} \mathbb{E} [p_l \nu].
\end{eqnarray}
Rewriting (\ref{gau06}), we obtain
\begin{eqnarray}
\mathbb{E} [p_l \nu] &\leq& q_n \Psi_{n} + (\Psi_{n} +1) e^{-\Psi_{n}}, \quad \forall \Psi_{n}>0.
\end{eqnarray}
Selecting $\Psi_{n} = \log (q_n^{-2})$ and defining $\varepsilon \triangleq nq_n$, we have
\begin{eqnarray}
 \bar{R}_g &\lesssim&  \frac{2 M \varepsilon}{N_0} \left( \log n -\log (\varepsilon^{-1}) \right).
\end{eqnarray}
As in the weak interference scenario we have $\varepsilon = o(1)$, it follows from the above equation that $\bar{R}_g = o (W \log n)$ in this scenario. Comparing with (\ref{gau8}), it follows that 
\begin{eqnarray} \label{gau9}
\bar{R}_g \lesssim \frac{W}{\hat{\alpha}} \log K. 
\end{eqnarray}
\textbf{Lower-bound} For the lower-bound, we consider the on-off power allocation scheme with $\tau_n = \log n -2 \log \log n$. Also, assume that $M=1$ (or equivalently, $n=K$). Noting $q_n = e^{-\tau_n}$, we obtain
\begin{eqnarray}
 \mathbb{E}[I_l]= (n-1) \hat{\alpha} q_n = \Theta (\log^2 n).
\end{eqnarray}
Therefore, using the result of Lemma 1, it is realized that with probability one $ (n-1) \hat{\alpha} q_n(1-\epsilon) \leq I_l \leq (n-1) \hat{\alpha} q_n (1+\epsilon) $, for some $\epsilon = o(1)$.  In other words, defining 
\begin{equation}
\varPhi (\nu) \triangleq \log \left( 1+\frac{p_l \nu}{(n-1) \hat{\alpha} q_n (1+\epsilon) +\frac{N_0 W}{M}} \right),
\end{equation}
it follows that
\begin{eqnarray}
\mathbb{P} \left\{ R_{l}(\textbf{\textrm{P}}^{(j)},\boldsymbol{\mathcal{L}}_{l}^{(j)}) < \varPhi (\nu)\right\} = o(1),
\end{eqnarray}
which implies that $R^*(\nu) \geq \varPhi (\nu)$. As a result,
\begin{eqnarray}
\bar{R}_g &\geq& nW \mathbb{E} [ \varPhi (\nu) ] \\
&=& nW \mathbb{E}  \left[ \log \left( 1+\frac{p_l \nu}{(n-1) \hat{\alpha} q_n (1+\epsilon) +\frac{N_0 W}{M}} \right) \right] \\
&\stackrel{(a)}{=}& nW \int_{\tau_n}^{\infty} \log \left( 1+\frac{\nu}{(n-1) \hat{\alpha} q_n (1+\epsilon) +\frac{N_0 W}{M}} \right) e^{-\nu} d \nu \\
&\geq& nW \int_{\tau_n}^{\Psi_{n}} \log \left( 1+\frac{\nu}{(n-1) \hat{\alpha} q_n (1+\epsilon) +\frac{N_0 W}{M}} \right) e^{-\nu} d \nu,
\end{eqnarray}
where $\Psi_{n} \triangleq \log n + 2 \log \log n$ and $(a)$ follows from the on-off power 
allocation assumption. As $(n-1) \hat{\alpha} q_n (1+\epsilon) = \Theta (\log^2 n)$, it 
follows that $\frac{\nu}{(n-1) \hat{\alpha} q_n (1+\epsilon) +\frac{N_0 W}{M}} = o(1)$ in the interval $[\tau_n,\Psi_n]$,
which implies that 
\begin{equation}
\log \left( 1+\frac{\nu}{(n-1) \hat{\alpha} q_n (1+\epsilon) +\frac{N_0 W}{M}} \right) \sim \frac{\nu}{(n-1) \hat{\alpha} q_n (1+\epsilon) +\frac{N_0 W}{M}},
\end{equation}
in the interval of integration $[\tau_n,\Psi_{n}]$. Hence,
\begin{eqnarray}
\bar{R}_g &\gtrsim& nW \int_{\tau_n}^{\Psi_{n}} \frac{\nu}{(n-1) \hat{\alpha} q_n (1+\epsilon) +\frac{N_0 W}{M}}  e^{-\nu} d \nu \\
&=& \frac{nW}{(n-1) \hat{\alpha} q_n (1+\epsilon) +\frac{N_0 W}{M}} \int_{\tau_n}^{\Psi_{n}} \nu e^{\nu} d \nu \\
&=& \frac{nW}{(n-1) \hat{\alpha} q_n (1+\epsilon) +\frac{N_0 W}{M}} \left( (\tau_n+1) e^{-\tau_n} - (\Psi_{n}+1)e^{-\Psi_{n}}\right) \\
&\stackrel{(a)}{\sim}&  \frac{nW \tau_n q_n}{(n-1) \hat{\alpha} q_n (1+\epsilon) +\frac{N_0 W}{M}} \\
&\sim& \frac{W}{\hat{\alpha}} \log n \\
&=& \frac{W}{\hat{\alpha}} \log K,
\end{eqnarray}
 where $(a)$ results from the facts that $(\Psi_{n}+1)e^{-\Psi_{n}} \ll (\tau_n+1) e^{-\tau_n}$ 
and $e^{-\tau_n} = q_n$. Combining the above equation with (\ref{gau9}), the proof of 
Theorem \ref{thmgau} follows.
\end{proof} 

\textbf{Remark 7-} Similar to the proof steps of Theorem 1, it can be shown that the 
optimum value of $M$ is equal to one. In fact, since the maximum guaranteed sum-rate 
of the network is achieved in the strong interference scenario in which the interference 
term scales as $n \hat{\alpha} q_n$ with probability one, it follows that the maximum 
network average sum-rate and the network guaranteed sum-rate are equal. Therefore, the optimum 
spectrum sharing for maximizing the network guaranteed sum-rate is the same as the one 
maximizing the average sum-rate of the network ($M=1$).

\section{Conclusion}\label{conclusion1}
In this paper, a distributed single-hop wireless network with $K$ links was considered, where 
the links were partitioned into a fixed number ($M$) of clusters each operating in a subchannel 
with bandwidth $\frac{W}{M}$. The subchannels were assumed to be orthogonal to each other. A general shadow-fading model, described by parameters $(\alpha,\varpi)$, was considered where 
$\alpha$ denotes the probability of shadowing and $\varpi$ ($\varpi \leq 1$)  represents the 
average cross-link gains. The maximum  achievable network throughput was studied in the 
asymptotic regime of $K \to \infty$. In the first part of the paper, the network throughput is 
defined as the \textit{average sum-rate} of the network, which is shown to scale as $\Theta 
(\log K)$. Moreover,  it was proved that in the strong interference scenario, the optimum power
allocation strategy for each user was a threshold-based on-off scheme. In the second part, the 
network throughput is defined as the \textit{guaranteed sum-rate}, when the outage probability 
approaches zero. In this scenario, it was demonstrated that the on-off power allocation scheme 
maximizes the network guaranteed sum-rate, which scales as $\frac{W}{\hat{\alpha}} \log K$.  
Moreover, the optimum spectrum sharing for maximizing the average sum-rate and guaranteed sum-rate 
is achieved at $M=1$.

\appendices
\section{Proof of Lemma \ref{lemma001}}\label{append01}

\renewcommand{\theequation}{A-\arabic{equation}}
\setcounter{equation}{0}

Let us define $\chi_{k} \triangleq \mathcal{L}_{ki}p_{k}$, where $\mathcal{L}_{ki}$
is independent of $p_{k}$, for $k \neq i$. Under a quasi-static Rayleigh fading channel
model, it is concluded that $\chi_{k}$'s are independent and identically distributed
(i.i.d.) random variables with
\begin{eqnarray}
\mathbb{E}\left[ \chi_{k} \right] & = & \mathbb{E}\left[ \mathcal{L}_{ki}p_{k} \right]=\hat{\alpha} q_{n}, \\
\textrm{Var} \left[ \chi_{k} \right] & = & \mathbb{E}\left[ \chi^{2}_{k} \right]-\mathbb{E}^{2}\left[ \chi_{k} \right]\\
&\stackrel{(a)}{\leq} & 2 \alpha \kappa q_{n}-(\hat{\alpha} q_{n})^{2},
\end{eqnarray}
where $\mathbb{E}\left[h_{ki}^{2} \right]=2$ and $\hat{\alpha} \triangleq \alpha \varpi$. Also, $(a)$ follows from the fact 
that $p_{k}^{2} \leq p_{k}$. Thus, $\mathbb{E}[p_{k}^{2}] \leq \mathbb{E}[p_{k}]=q_{n}$. The
interference $I_{i} = \sum_{\substack{k  \in \mathbb{C}_{j} \\ k\neq i}}\chi_{k}$ is a random variable with mean $\mu_{n}$ and variance $\vartheta_{n}^{2}$, where
\begin{eqnarray}
\mu_{n} & \triangleq & \mathbb{E}\left[ I_i \right]=(n-1)\hat{\alpha} q_{n},\\
\vartheta_{n}^{2} & \triangleq & \textrm{Var} \left[I_i \right] \leq (n-1)(2\alpha \kappa q_{n}-(\hat{\alpha}q_{n})^{2}) \leq (n-1)(2\alpha \kappa q_{n}).
\end{eqnarray}
Using the \textit{Central Limit Theorem} \cite[p. 183]{Petrovbook95}, we obtain
\begin{eqnarray}
\mathbb{P} \lbrace \vert I_{i}-\mu_{n} \vert <\psi_{n} \rbrace  &\approx&  1-Q \left(\dfrac{\psi_{n}}{\vartheta_{n}} \right) \\
&\stackrel{(a)}{\geq}& 1-e^{-\frac{\psi^2_{n}}{2 \vartheta^2_{n}}},
\end{eqnarray}
for all $\psi_{n} >0$ such that $\psi_n = o \left( n^{\frac{1}{6}} \vartheta_n\right)$. In the 
above equation, the $Q(.)$ function is defined as $Q(x)\triangleq \frac{1}{\sqrt{2\pi}}\int_{x}^{\infty}e^{-u^{2}/2}du$, and $(a)$ follows from the fact that $Q(x) \leq e^{-\frac{x^2}{2}}$, $\forall x> 0$. Selecting $\psi_n = \left( n q_n\right)^{\frac{1}{8}} \sqrt{2} \vartheta_n$, we obtain
\begin{eqnarray}
 \mathbb{P} \lbrace \vert I_{i}-\mu_{n} \vert <\psi_{n} \rbrace &\geq& 1-e^{-\left( n q_n\right)^{\frac{1}{4}}}.
\end{eqnarray}
Therefore, defining $\varepsilon \triangleq \frac{\psi_n}{\mu_n} = O \left( (nq_n)^{-\frac{3}{8}} \right)$, we have
\begin{eqnarray}
 \mathbb{P} \lbrace \mu_n \left( 1-\varepsilon\right) \leq I_i \leq \mu_n \left( 1 + \varepsilon\right)\rbrace &\geq& 1-e^{-\left( n q_n\right)^{\frac{1}{4}}}.
\end{eqnarray}
Noting that $n q_n \to \infty$, it follows that $I_i \sim \mu_n$, with probability one. 
Now, we show a stronger statement, which is, the contribution of the realizations in 
which $|I_i - \mu_n| > \psi_n$ in the network average sum-rate is negligible. For this purpose, 
we give a lower-bound and an upper-bound for the network average sum-rate and show that these 
bounds converge to each other in the strong interference regime, when $n q_n \to \infty$. 
A lower-bound denoted by $\bar{R}_{ave}^{(L)}$, can be given by 
\begin{eqnarray}
\bar{R}_{ave}^{(L)} &\triangleq& nW \mathbb{E} \left.\left[ \log \left( 1+ \frac{\hat{p}_i h_{ii}}{I_i + \frac{N_0 W}{M}}\right) \right|  \vert I_{i}-\mu_{n} \vert <\psi_{n}\right] \mathbb{P} \{ \vert I_{i}-\mu_{n} \vert <\psi_{n}\} \\
&\geq& nW \mathbb{E} \left[ \log \left( 1+ \frac{\hat{p}_i h_{ii}}{\mu_n (1+ \varepsilon) + \frac{N_0 W}{M}}\right) \right] \left[ 1-e^{-\left( n q_n\right)^{\frac{1}{4}}} \right],
\end{eqnarray}
which scales as $\frac{W}{\hat{\alpha}} \log n$ (as shown in the proof of Theorem 1, by 
optimizing the power allocation function). An upper-bound for the network average sum-rate, 
denoted by $\bar{R}_{ave}^{(U)}$, can be given as
\begin{eqnarray}
\bar{R}_{ave}^{(U)} &=& nW \mathbb{E} \left.\left[ \log \left( 1+ \frac{\hat{p}_i h_{ii}}{I_i + \frac{N_0 W}{M}}\right) \right|  \vert I_{i}-\mu_{n} \vert <\psi_{n}\right] \mathbb{P} \{ \vert I_{i}-\mu_{n} \vert <\psi_{n}\} + \notag\\
&& nW \mathbb{E} \left.\left[ \log \left( 1+ \frac{\hat{p}_i h_{ii}}{I_i + \frac{N_0 W}{M}}\right) \right|  \vert I_{i}-\mu_{n} \vert \geq \psi_{n}\right] \mathbb{P} \{ \vert I_{i}-\mu_{n} \vert \geq \psi_{n}\} \\
&\leq& \bar{R}_{ave}^{(L)} + nW\mathbb{E} \left[ \log \left( 1+ \frac{\hat{p}_i h_{ii}}{\frac{N_0 W}{M}}\right) \right] e^{-\left( n q_n\right)^{\frac{1}{4}}}\\
&\stackrel{(a)}{\leq}& \bar{R}_{ave}^{(L)} + n W \mathbb{E} \left[  \dfrac{\hat{p}_i h_{ii}}{\frac{N_0 W}{M}} \right] e^{-\left( n q_n\right)^{\frac{1}{4}}}\\
&\stackrel{(b)}{=}& \bar{R}_{ave}^{(L)}  + W O (nq_n \log n) e^{-\left( n q_n\right)^{\frac{1}{4}}} \\
&\stackrel{(c)}{\sim}& \bar{R}_{ave}^{(L)}.
\end{eqnarray}
In the above equation, $(a)$ follows from the fact that $\log (1+x) \leq x$, $(b)$ comes from the facts 
that $\mathbb{E} \{p_i h_{ii}\} \lesssim  q_n \log n$ (this is shown in the proof of Theorem \ref{thmgau}) 
and $\frac{N_{0}W}{M}$ is fixed, and 
finally, $(c)$ results from the fact that as $n q_n \to \infty$, $nq_n e^{-(nq_n)^{\frac{1}{4}}} \to 0$. 
The above equation implies that substituting $I_i$ by its mean ($(n-1)\hat{\alpha} q_n$) does not affect 
the analysis of the network average sum-rate in the asymptotic case of $K \to \infty$.

\section{Proof of Lemma \ref{lemma002}}\label{append02}

\renewcommand{\theequation}{B-\arabic{equation}}
\setcounter{equation}{0}

Denoting $\mathbb{T}_{j} \triangleq \lbrace l \in
\mathbb{C}_{j}~\vert~h_{ll} > h_{Th} \rbrace$, the cardinality of
the set $\mathbb{T}_{j}$ is a binomial random variable with the mean
$n\mathbb{P}\lbrace h_{ll} > h_{Th}\rbrace$. From (\ref{eqn: 05}),
we have
\begin{equation} \label{rsum1}
\bar{R}_{ave} = \sum_{j=1}^{M} \mathbb{E}\left[ \sum_{l \in \mathbb{C}_{j}}R_{l}(\hat{\textbf{\textrm{P}}}^{(j)},\boldsymbol{\mathcal{L}}_{l}^{(j)}) \right],
\end{equation}
where 
\begin{equation}\label{sd9}
\mathbb{E} \left[ \sum_{l \in \mathbb{C}_{j}}R_{l}(\hat{\textbf{\textrm{P}}}^{(j)},\boldsymbol{\mathcal{L}}_{l}^{(j)}) \right]=
\mathbb{E} \left[ \sum_{l \in \mathbb{T}_{j}}R_{l}(\hat{\textbf{\textrm{P}}}^{(j)},\boldsymbol{\mathcal{L}}_{l}^{(j)}) \right]+\mathbb{E} \left[ \sum_{l \in \mathbb{T}^{C}_{j}}R_{l}(\hat{\textbf{\textrm{P}}}^{(j)},\boldsymbol{\mathcal{L}}_{l}^{(j)}) \right],
\end{equation}
in which $\mathbb{T}_{j}^C$ denotes the complement
of $\mathbb{T}_{j}$. Note that
\begin{eqnarray}
\mathbb{E} \left[ \sum_{l \in \mathbb{T}_{j}}R_{l}(\hat{\textbf{\textrm{P}}}^{(j)},\boldsymbol{\mathcal{L}}_{l}^{(j)}) \right]& = & n \frac{W}{M}\mathbb{E} \left. \left[\log \left (1+ \dfrac{h_{ll}\hat{p}_{l}}{I_{l}+\frac{N_{0}W}{M}}\right) \right | h_{ll}>h_{Th} \right]\mathbb{P}\lbrace h_{ll} > h_{Th} \rbrace\\
& \leq & n \frac{W}{M} \mathbb{E} \left. \left[\log \left (1+ \dfrac{h_{ll}}{\frac{N_{0}W}{M}}\right) \right | h_{ll}>h_{Th} \right]\mathbb{P}\lbrace h_{ll} > h_{Th} \rbrace\\
&\stackrel{(a)}{\leq} & \dfrac{n}{N_{0}} e^{-h_{Th}}\mathbb{E} \left[h_{ll} | h_{ll}>h_{Th} \right]\\
\label{dfs} &= & \dfrac{n}{N_{0}} e^{-h_{Th}} (1+h_{Th}),
\end{eqnarray}
where $(a)$ follows from $\log (1+x) \leq x$, for $x > 0$. It is
observed that for $h_{Th}=c\log n$, where $c >1$, the right hand
side of (\ref{dfs}) tends to zero as $n \rightarrow \infty$. Thus,
\begin{equation}
\lim_{n \rightarrow \infty } \mathbb{E} \left[ \sum_{l \in \mathbb{T}_{j}}R_{l}(\hat{\textbf{\textrm{P}}}^{(j)},\boldsymbol{\mathcal{L}}_{l}^{(j)}) \right]=0.
\end{equation}
Consequently,
\begin{equation}
\lim_{n \rightarrow \infty } \sum_{j=1}^{M}\mathbb{E} \left[ \sum_{l \in \mathbb{T}_{j}}R_{l}(\hat{\textbf{\textrm{P}}}^{(j)},\boldsymbol{\mathcal{L}}_{l}^{(j)}) \right]=0,
\end{equation}
and this completes the proof of the lemma.

\section{Proof of Lemma \ref{lemma003}}\label{append03}

\renewcommand{\theequation}{C-\arabic{equation}}
\setcounter{equation}{0}

Using (\ref{asy01}), we have
\begin{eqnarray}
\notag \mathbb{E}[\Xi_{i}(\hat{p}_{i},h_{ii})]&\approx&\dfrac{W}{M\lambda} \mathbb{E}[h_{ii}\hat{p}_{i}] + n\dfrac{\alpha W \mu}{M\lambda'} \left( 1-\dfrac{\varpi }{\lambda'}\mathbb{E}[\hat{p}_{i}]+ \dfrac{2\kappa}{\lambda'^2}\mathbb{E}[\hat{p}_{i}^{2}]- \dfrac{6\eta}{\lambda'^3}\mathbb{E}[\hat{p}_{i}^{3}]\right)+\\
&& n(1-\alpha)\dfrac{W \mu}{M\lambda'}\\
\notag & \stackrel{(a)}{=} &\dfrac{W}{M\lambda}(1+\tau_{n})q_{n}-\dfrac{n \hat{\alpha}W}{M\lambda'^2}(1+\tau_{n})q_{n}^{2}+\dfrac{n \alpha W2\kappa}{M\lambda'^3}(1+\tau_{n})q_{n}^{2}-\\
&& \dfrac{n\alpha W6 \eta}{M\lambda'^4}(1+\tau_{n})q_{n}^{2} +\dfrac{nW}{M\lambda'}(1+\tau_{n})q_{n}\\
\label{uppera1} & \stackrel{(b)}{\approx} & \dfrac{W}{M\hat{\alpha}} \left( 1+\tau_{n}+\dfrac{\xi_{1}}{n^{2}}(1+\tau_{n})e^{\tau_{n}}-\dfrac{\xi_{2}}{n^3}(1+\tau_{n})e^{2\tau_{n}} \right),
\end{eqnarray}
where $\xi_{1} \triangleq \dfrac{2\kappa}{\varpi \hat{\alpha}}$ and
$\xi_2 \triangleq \dfrac{6\eta}{\varpi \hat{\alpha}^2}$. In the
above equation, $(a)$ follows from the fact that
$\mathbb{E}[h_{ii}\hat{p}_{i}]=\mu=(1+\tau_{n})q_{n}$, and $(b)$
results from i)~$\lambda = (n-1) \hat{\alpha} q_n +
\frac{N_{0}W}{M} \approx n \hat{\alpha} q_n$ and $\lambda' \approx n \hat{\alpha} q_n$ incurred by  the fact that $\lambda \gg 1$, and
ii)~$q_n=e^{-\tau_n}$. Since $n \hat{\alpha} q_n \to \infty $, it follows that the right hand side of
(\ref{uppera1}) is a monotonically increasing function of $\tau_n$, which attains its maximum when $\tau_n$ takes its maximum feasible value. The maximum feasible value of $\tau_n$, denoted as $\hat{\tau}_n$, can be obtained as
\begin{eqnarray}
 n \hat{\alpha} e^{-\tau_n} \to \infty  \Longrightarrow \hat{\tau}_n \sim \log n.
\end{eqnarray}
Thus, the maximum achievable value for $\mathbb{E}[\Xi_{i}(\hat{p}_{i},h_{ii})]$ scales as $\dfrac{W}{M\hat{\alpha}} \log n $.

\section{Proof of Lemma \ref{lemma004}}\label{append04}

\renewcommand{\theequation}{D-\arabic{equation}}
\setcounter{equation}{0} $i)$ Using (\ref{utility1}) and assuming
that all users follow the on-off power allocation policy,
$\mathbb{E}[u_{i}(\hat{p}_{i},h_{ii})]$ can be expressed as
\begin{equation}\label{averutility}
\mathbb{E}[u_{i}(\hat{p}_{i},h_{ii})] = \sum_{l \in \mathbb{C}_{j}} \mathbb{E}\left[R_{l}(\hat{\textbf{\textrm{P}}}^{(j)},\boldsymbol{\mathcal{L}}_{l}^{(j)})\right],~~j=1,...,M,
\end{equation}
where the expectation is computed with respect to $h_{ll}$ and
$I_{l}$. Noting that $q_{n}=\mathbb{P} \left\lbrace h_{ll} >
\tau_{n}\right\rbrace$, we have
\begin{eqnarray}
\nonumber \mathbb{E}\left[R_{l}(\hat{\textbf{\textrm{P}}}^{(j)},\boldsymbol{\mathcal{L}}_{l}^{(j)})\right] & = & \mathbb{E}\left. \left[R_{l}(\hat{\textbf{\textrm{P}}}^{(j)},\boldsymbol{\mathcal{L}}_{l}^{(j)}) \right | h_{ll}>\tau_{n} \right]\mathbb{P} \left\lbrace h_{ll} > \tau_{n}\right\rbrace+\\
&&\mathbb{E}\left. \left[R_{l}(\hat{\textbf{\textrm{P}}}^{(j)},\boldsymbol{\mathcal{L}}_{l}^{(j)}) \right | h_{ll}\leq \tau_{n} \right]\mathbb{P} \left\lbrace h_{ll} \leq \tau_{n}\right\rbrace\\
\nonumber & = & q_{n} \mathbb{E}\left. \left[R_{l}(\hat{\textbf{\textrm{P}}}^{(j)},\boldsymbol{\mathcal{L}}_{l}^{(j)}) \right | h_{ll}>\tau_{n} \right] + (1-q_{n})\mathbb{E} \left. \left[R_{l}(\hat{\textbf{\textrm{P}}}^{(j)},\boldsymbol{\mathcal{L}}_{l}^{(j)}) \right | h_{ll}\leq \tau_{n} \right].
\end{eqnarray}
Since for $h_{ll} \leq \tau_{n}$, $\hat{p}_{l}=0$, it is concluded that
\begin{equation} \label {eqn:q9}
\mathbb{E}\left[R_{l}(\hat{\textbf{\textrm{P}}}^{(j)},\boldsymbol{\mathcal{L}}_{l}^{(j)})\right]=\dfrac{q_{n}W}{M}\mathbb{E} \left. \left[\log \left( 1+\dfrac{h_{ll}}{I_{l}+\frac{N_{0}W}{M}}\right)  \right | h_{ll}>\tau_{n} \right].
\end{equation}
For large values of $K$, we can apply Lemma \ref{lemma001} to obtain
\begin{eqnarray}
\mathbb{E}\left[R_{l}(\hat{\textbf{\textrm{P}}}^{(j)},\boldsymbol{\mathcal{L}}_{l}^{(j)})\right] &\approx & \dfrac{q_{n}W}{M}\mathbb{E} \left. \left[  \log \left( 1+\dfrac{h_{ll}}{(n-1)\hat{\alpha} q_{n}+\frac{N_{0}W}{M}}\right) \right | h_{ll}>\tau_{n} \right] \\
\label {eqn:q10} &=& \dfrac{q_{n}W}{M}\mathbb{E} \left. \left[  \log \left( 1+\dfrac{h_{ll}}{\lambda}\right) \right | h_{ll}>\tau_{n} \right],
\end{eqnarray}
where the expectation is computed with respect to $h_{ll}$. 
%Here, we assume that $\lambda \rightarrow \infty$ when $n \rightarrow \infty$ in order to use the approximation $\log(1+z)\approx
%z-\dfrac{z^{2}}{2}$ for $\vert z \vert \ll 1$. Again, as we will see
%later, this assumption is feasible for the optimum value of $q_{n}$.
%To this end, (\ref{eqn:q10}) can be written as
Using the Taylor series for  $\log(1+x)$, (\ref{eqn:q10}) can be written as
\begin{eqnarray}\label {eqn:q101}
\mathbb{E}\left[R_{l}(\hat{\textbf{\textrm{P}}}^{(j)},\boldsymbol{\mathcal{L}}_{l}^{(j)})\right] &\approx& \dfrac{q_{n}W}{M} \sum_{k=1}^{\infty}\dfrac{(-1)^{k-1}}{k \lambda^k} \mathbb{E}\left. \left[  h_{ll}^k \right | h_{ll}>\tau_{n} \right]  \\
&\stackrel{(a)}{\approx}& \dfrac{q_{n}W}{M} \sum_{k=1}^{\infty}\dfrac{(-1)^{k-1}}{k (n\hat{\alpha} q_{n})^k} \mathbb{E}\left. \left[  h_{ll}^k \right | h_{ll}>\tau_{n} \right] \\
\label{opio}& \stackrel{(b)}{\approx} & \dfrac{q_{n}W}{M} \sum_{k=1}^{\infty}\dfrac{(-1)^{k-1} \tau_n^k}{k (n\hat{\alpha} q_{n})^k} \\
&=& \dfrac{q_{n}W}{M} \log \left( 1+\frac{\tau_n}{n \hat{\alpha} q_n}\right) \\
&\stackrel{(c)}{=}& \dfrac{e^{-\tau_n}W}{M} \log \left( 1+\frac{\tau_n e^{\tau_n}}{n \hat{\alpha} }\right),
\end{eqnarray}
where $(a)$ follows from the fact that for large values of $n$, $\lambda \approx n \hat{\alpha} q_n$. 
Also, $(b)$ results from the fact that under a Rayleigh fading channel model,
\begin{equation}
\mathbb{E} \left. \left[  h_{ll} \right | h_{ll}>\tau_{n} \right]=1+\tau_{n}, 
\end{equation}
\begin{equation}
\mathbb{E} \left. \left[  h_{ll}^{k} \right | h_{ll}>\tau_{n} \right]=\tau_{n}^{k}+k \mathbb{E} \left. \left[  h_{ll}^{k-1} \right | h_{ll}>\tau_{n} \right]. 
\end{equation}
Since $\lambda \gg 1$, the term $\frac{\mathbb{E} \left. \left[  h_{ll}^{k-1} \right | h_{ll}>\tau_{n} \right]}{\lambda^k} \ll \frac{\mathbb{E} \left. \left[  h_{ll}^{k-1} \right | h_{ll}>\tau_{n} \right]}{\lambda^{k-1}}$, which implies that we can neglect this term and simply write $\mathbb{E} \left. \left[  h_{ll}^{k} \right | h_{ll}>\tau_{n} \right] \approx \tau_n^k$. $(c)$ results from $q_n = e^{-\tau_n}$. Thus, (\ref{averutility}) can be simplified as
\begin{equation}\label{aveutility1}
\mathbb{E}[u_{i}(\hat{p}_{i},h_{ii})] \approx \dfrac{ne^{-\tau_n}W}{M} \log \left( 1+\frac{\tau_n e^{\tau_n}}{n \hat{\alpha} }\right).
\end{equation}
In order to find the optimum threshold value:
\begin{equation} \label{opti87}
\hat{\tau}_{n}=\arg~\max_{\tau_{n}}~ \mathbb{E}[u_{i}(\hat{p}_{i},h_{ii})], 
\end{equation}
we set the derivative of the right hand side of (\ref{aveutility1}) with respect to $\tau_n$ to zero:
\begin{eqnarray} \label{eqn:d9}
-e^{-\hat{\tau_n}} \log \left( 1+\frac{\hat{\tau}_n e^{\hat{\tau}_n}}{n \hat{\alpha}}\right) + \frac{1+\hat{\tau}_n}{n \hat{\alpha} + \hat{\tau}_n e^{\hat{\tau}_n}} =0,
\end{eqnarray}
which after some manipulations yields
\begin{eqnarray} \label{eqn: 132}
 \hat{\tau}_n = \log n - 2 \log \log n + O(1).
\end{eqnarray}

$ii)$ Using (\ref{eqn: 132}), it is concluded that
\begin{eqnarray}
q_{n}  &=& e^{-\tau_{n}}\\
&=& \delta \dfrac{\log ^{2} n }{n},
\end{eqnarray}
where $\delta$ is a constant. 

$iii)$ Using  (\ref{eqn: 132}),  we have
\begin{eqnarray}
 \frac{\hat{\tau}_n e^{\hat{\tau}_n}}{n \hat{\alpha}} = \Theta \left( \frac{1}{\log n}\right),
\end{eqnarray}
which implies that the right hand side of (\ref{aveutility1}) can be written as
\begin{eqnarray} \label{eqn:q102}
 \mbox{R\!H (\ref{aveutility1})} &\approx& \frac{W \hat{\tau}_n}{M \hat{\alpha}}.  
\end{eqnarray}
Thus, the maximum value for $\mathbb{E}[u_{i}(\hat{p}_{i},h_{ii})]$ in (\ref{aveutility1}) 
scales as $\dfrac{W}{M\hat{\alpha}} \log n $.

\section{Proof of Theorem \ref{corol004}}\label{append05}

\renewcommand{\theequation}{E-\arabic{equation}}
\setcounter{equation}{0}

Let us define $\mathbb{A}_{j}$ as the set of active links in cluster
$j$. The random variable $m_{j}$ denotes the cardinality of the set
$\mathbb{A}_{j}$. Noting that for $M = \Theta(K) $, $\lim_{K \to
\infty}  \frac{M}{K} $ is constant, it is concluded that $n$ and
$m_{j} \in [1,n]$ do not grow with $K$. To obtain the network
average sum-rate, we assume that among $M$ clusters, $\Gamma$ clusters
have $m_{j}=1$ and the rest have $m_{j}>1$. We first obtain an upper
bound on the average sum-rate in each cluster when $m_{j}=1$, $1\leq j
\leq M$. Clearly, since only one user in each cluster activates its
transmitter, $I_{i}=0$. Thus, by using (\ref{fds}), the maximum
achievable average sum-rate of cluster $\mathbb{C}_{j}$ is computed as
\begin{eqnarray}
\label{eq01} \bar{R}_{ave}^{(j)} = \dfrac{W}{M} \mathbb{E}\left[ \log \left(1+\dfrac{M}{N_{0}W}h_{\max}\right) \right],
\end{eqnarray}
where $h_{\max} \triangleq \max~ \lbrace h_{ii}\rbrace_{_{i \in
\mathbb{C}_{j}}}$ is a random variable. Since $\log x$ is a concave
function of $x$, an upper bound of (\ref{eq01}) is obtained through
\textit{Jensen's inequality}, $\mathbb{E}\left[\log x
\right]\leq\log (\mathbb{E}\left[x \right])$, $x>0$. Thus,
\begin{equation} \label{eq02}
\bar{R}_{ave}^{(j)} \leq \dfrac{W}{M}  \log \left(1+\dfrac{M}{N_{0}W} \mathbb{E}\left[ h_{\max} \right] \right).
\end{equation}
Under a Rayleigh fading channel model and noting that $\lbrace
h_{ii}\rbrace$ is a set of i.i.d. random variables over $i \in
\mathbb{C}_{j}$, we have
\begin{eqnarray}
F_{h_{_{max}}}(y) & = & \mathbb{P}\lbrace h_{\max} \leq y \rbrace,  ~~y>0 \\
& = & \prod _{i \in \mathbb{C}_{j}} \mathbb{P} \lbrace h_{ii} \leq y \rbrace \\
& = & \left( 1-e^{-y} \right)^{n},
\end{eqnarray}
where $F_{h_{_{max}}}(.)$ is the cumulative distribution function
(CDF) of $h_{\max}$. Hence,
\begin{equation}
\mathbb{E}\left[ h_{\max} \right]  = \int_{0}^{\infty} n y e^{-y}\left( 1-e^{-y} \right)^{n-1} dy.
\end{equation}
Since  $\left( 1-e^{-y} \right)^{n-1}\leq 1$, we arrive at the following inequality
\begin{equation}\label{eq03}
 \mathbb{E}\left[ h_{\max} \right] \leq  \int_{0}^{\infty} n y e^{-y} dy =n.
\end{equation}
Consequently, the upper bound of (\ref{eq02}) can be simplified as
\begin{equation} \label{eq04}
\bar{R}_{ave}^{(j)} \leq \dfrac{W}{M}  \log \left(1+\dfrac{K}{N_{0}W} \right).
\end{equation}

For $m_{j} >1$ and due to the shadowing effect with parameters
$(\alpha,\varpi)$, the average sum-rate of cluster $\mathbb{C}_{j}$ can be written as
\begin{equation}
\label{eqn 35} \bar{R}_{ave}^{(j)} =  \sum_{i \in \mathbb{A}_{j}}\dfrac{W}{M} \mathbb{E}\left[ \log \left(1+\dfrac{h_{ii}}{\sum_{\begin{subarray}{c}
k \in \mathbb{A}_{j} \\
k\neq i \end{subarray}}u _{k}\beta_{ki} h_{ki}+\frac{N_{0}W}{M}}\right) \right],
\end{equation}
where $u _{k}$'s are Bernoulli random variables with parameter $\alpha$. Thus,
\begin{eqnarray}
\bar{R}_{ave}^{(j)} & = &   \dfrac{W}{M} \sum_{i \in \mathbb{A}_{j}} \sum_{l=0}^{m_{j}-1} \binom{m_{j}-1}{l} \alpha ^{l} (1-\alpha)^{m_{j}-1-l}\mathbb{E}\left[ \log \left(1+\dfrac{h_{ii}}{\Sigma_{l}+ \frac{N_{0}W}{M}}\right) \right]\\
\nonumber & = & \dfrac{W}{M} \sum_{i \in \mathbb{A}_{j}} (1-\alpha)^{m_{j}-1}\mathbb{E}\left[ \log \left(1+\dfrac{h_{ii}}{ \frac{N_{0}W}{M}}\right) \right]+\\
\label{dd} & & \dfrac{W}{M} \sum_{i \in \mathbb{A}_{j}} \sum_{l=1}^{m_{j}-1} \binom{m_{j}-1}{l} \alpha ^{l} (1-\alpha)^{m_{j}-1-l}\mathbb{E}\left[ \log \left(1+\dfrac{h_{ii}}{\Sigma_{l}+ \frac{N_{0}W}{M}}\right) \right],
\end{eqnarray}
where $\Sigma_{l}$ is the sum of $l$ i.i.d random variables
$\{Z_i\}_{i=1}^l$, where $Z_i \triangleq \beta_{ki} h_{ki}$, $k \neq
i$. For $m_{j}>1$, $\Sigma_{l}$ is greater than the interference
term caused by one interfering link. Thus, an upper bound on the
average sum-rate of cluster $\mathbb{C}_{j}$ is computed as
\begin{eqnarray}
\nonumber \bar{R}_{ave}^{(j)}  & \leq &   \dfrac{W}{M} m_{j} (1-\alpha)^{m_{j}-1}\mathbb{E}\left[ \log \left(1+\dfrac{Y}{ \frac{N_{0}W}{M}}\right) \right]+\\
\label{eqn 305} && \dfrac{W}{M} \sum_{i \in \mathbb{A}_{j}} \sum_{l=1}^{m_{j}-1} \binom{m_{j}-1}{l} \alpha ^{l} (1-\alpha)^{m_{j}-1-l}\mathbb{E}\left[ \log \left(1+\dfrac{Y}{Z_{i}+\frac{N_{0}W}{M}}\right) \right],
\end{eqnarray}
where $Y \triangleq h_{\max}=\max~ \lbrace h_{ii}\rbrace_{_{i \in \mathbb{C}_{j}}}$. According to binomial formula, we have
\begin{equation}
\sum_{l=1}^{m_{j}-1} \binom{m_{j}-1}{l} \alpha ^{l} (1-\alpha)^{m_{j}-1-l} =1-(1-\alpha)^{m_{j}-1}.
\end{equation}
Thus,
\begin{eqnarray}
\nonumber  \bar{R}_{ave}^{(j)}  & \leq &  \dfrac{W}{M} m_{j} (1-\alpha)^{m_{j}-1}\mathbb{E}\left[ \log \left(1+\dfrac{Y}{ \frac{N_{0}W}{M}}\right) \right]+\\
\label{ef2} && \dfrac{W}{M} m_{j} \left (1-(1-\alpha)^{m_{j}-1}\right )\mathbb{E}\left[ \log \left(1+\dfrac{Y}{\beta_{ki} h_{ki}+\frac{N_{0}W}{M}}\right) \right].
\end{eqnarray}
We have
\begin{equation}\label{beta_th}
\mathbb{E}\left[ \log \left(1+\dfrac{Y}{\beta_{ki} h_{ki}+\frac{N_{0}W}{M}}\right) \right] \leq  \mathbb{E}  \left[ \log \left(1+\dfrac{Y}{\beta_{\min} h_{ki}}\right) \right].
\end{equation}
 Defining $Z
\triangleq \beta_{\min} h_{ki}$ and $X \triangleq \frac{Y}{Z}$, the
CDF of $X$ can be evaluated as
\begin{eqnarray}
F_{X}(x) & = & \mathbb{P}\lbrace X \leq x \rbrace,~~~x>0 \\
& = & \mathbb{P} \lbrace Y \leq Zx \rbrace \\
& = & \int_{0}^{\infty} \mathbb{P} \lbrace Y \leq Zx \vert Z =z \rbrace  f_{Z}(z)dz\\
& = & \int_{0}^{\infty} \left( 1-e^{-zx} \right)^{n} \frac{1}{\beta_{\min}}e^{-\frac{z}{\beta_{\min}}} dz\\
& = & \int_{0}^{\infty} \left( 1-e^{-t \beta_{\min} x} \right)^{n} e^{-t} dt.
\end{eqnarray}
Thus, the probability density function of $X$ can be written as
\begin{eqnarray}
f_{X}(x) & = & \dfrac{d F_{X} (x)}{dx} \\
& = & \beta_{\min} \int_{0}^{\infty} nt e^{-t(1+ \beta_{\min} x)}\left( 1-e^{-t \beta_{\min} x} \right)^{n-1} dt\\
\label{eq06} & \leq & \beta_{\min} \int_{0}^{\infty} nt e^{-t(1+ \beta_{\min} x)} dt \\
&=& \dfrac{n \beta_{\min}}{(1+\beta_{\min} x)^{2}}.
\end{eqnarray}
Using the above equation, the right hand side of (\ref{beta_th}) can be upper-bounded as
\begin{eqnarray}
\mathbb{E}  \left[ \log \left(1+\dfrac{Y}{\beta_{\min} h_{ki} }\right) \right] &=& \int_{0}^{\infty} f_{X}(x) \log (1+x) dx \\
&\leq& n \beta_{\min} \int_{0}^{\infty} \frac{\log(1+x)}{(1+\beta_{\min} x)^{2}} dx \\
&=& \frac{-n \log \beta_{\min}}{1-\beta_{\min}} \\
&=& \Theta (1),
\end{eqnarray}
where the last line follows from the fact that $0< \beta_{\min} \leq 1$.
Substituting the above equation in (\ref{ef2}) yields
\begin{eqnarray} \label{yy}
\bar{R}_{ave}^{(j)}  & \leq &  \dfrac{W}{M} m_{j} (1-\alpha)^{m_{j}-1}\mathbb{E}\left[ \log \left(1+\dfrac{Y}{ \frac{N_{0}W}{M}}\right) \right]+\notag\\
&& \dfrac{W}{M} m_{j} \left (1-(1-\alpha)^{m_{j}-1}\right) \Theta(1) \\
&\stackrel{(a)}{\leq}&  \dfrac{W}{M} m_{j} (1-\alpha)^{m_{j}-1} \log \left(1+\dfrac{K}{N_{0}W} \right) + \Theta \left(\frac{W}{M}\right) \\
&=& \dfrac{W}{M} m_{j} (1-\alpha)^{m_{j}-1} \log \left(1+\dfrac{K}{N_{0}W} \right) \left[1+o(1) \right],
\end{eqnarray}
where $(a)$ follows from (\ref{eq04}) and the fact that $m_j \in \lbrace 2,...,n \rbrace$ does not scale with $K$.

Let us assume that among $M$ clusters, $\Gamma$ clusters have
$m_{j}=1$ and for the $M-\Gamma$ of the rest, the number of active
links in each cluster is greater than one. By using (\ref{eq04}) and
(\ref{yy}), an upper bound on the network average sum-rate is obtained as
\begin{eqnarray}
\nonumber \bar{R}_{ave} &\leq &  \dfrac{\Gamma W}{M} \log \left(1+\dfrac{K}{N_{0}W} \right)+\\
\label{tr1} && (M-\Gamma) \dfrac{W}{M} m_{j} (1-\alpha)^{m_{j}-1} \log \left(1+\dfrac{K}{N_{0}W} \right) \left[1+o(1) \right].
\end{eqnarray}
To compare this upper-bounded with the computed network average sum-rate in the case of $M=1$, we note that as $\varpi \leq 1$
and $\alpha < 1$, we have $\hat{\alpha} < 1$ and consequently,
\begin{equation}
\dfrac{\Gamma W}{M} \log \left(1+\dfrac{K}{N_{0}W} \right) <  \dfrac{\Gamma W}{M \hat{\alpha}} \log \left(1+\dfrac{K}{N_{0}W} \right).
\end{equation}
To prove that the maximum network average sum-rate obtained in
(\ref{tr1}) is less than that value obtained for $M=1$ from
(\ref{er}), it is sufficient to show
\begin{equation} \label{up2}
(M-\Gamma) \dfrac{W}{M} m_{j} (1-\alpha)^{m_{j}-1} \log \left(1+\dfrac{K}{N_{0}W} \right)  <  (M-\Gamma)\dfrac{W}{M \hat{\alpha}} \log \left(1+\dfrac{K}{N_{0}W} \right),
\end{equation}
or
\begin{equation}
m_{j}(1-\alpha)^{m_{j}-1} < \frac{1}{\hat{\alpha}}.
\end{equation}
Since $\hat{\alpha} \leq \alpha$, it is sufficient to show that $m_{j}(1-\alpha)^{m_{j}-1} < \frac{1}{\alpha}$.
Defining $\Lambda(\alpha)=\alpha m_{j} (1-\alpha)^{m_{j}-1}$, we have
\begin{equation}
\dfrac{\partial \Lambda(\alpha)}{\partial \alpha} = m_{j}(1-\alpha)^{m_{j}-2} (1- \alpha m_{j}).
\end{equation}
Thus, the extremum points of $\Lambda(\alpha)$ are located at $\alpha = 1$
and $\alpha=\frac{1}{m_{j}}$, where $m_j \in \lbrace 2,...,n \rbrace$.
It is observed that
\begin{equation}
\Lambda(1) = 0 <1,
\end{equation}
and
\begin{equation}
\Lambda \left(\dfrac{1}{m_{j}}\right) = \left( \dfrac{m_{j}-1}{m_{j}} \right)^{m_{j}-1} <1.
\end{equation}
Since $\Lambda(\alpha) < 1$, we conclude (\ref{up2}), which implies that
the maximum average sum-rate of the network for $M = \Theta(K)$ is less than
that of $M=1$. Knowing the fact that for $M = o(K)$, similar to the result of 
Theorem 1, one can show that the maximum average sum-rate of the network is achieved 
at $M=1$, it is concluded that using the on-off allocation scheme, the maximum 
average sum-rate of the network is achieved at $M=1$, for all values of $1 \leq M \leq K$.

\section{Proof of Remark 4}\label{append06}

\renewcommand{\theequation}{F-\arabic{equation}}
\setcounter{equation}{0}

Using (\ref{eqn: 02}) and (\ref{eqn: 05}) and for every value of $1 \leq M \leq K$ and $\alpha = 0$,
the average sum-rate of the network is simplified as
\begin{equation}\label{rr0}
\bar{R}_{ave}  =  \sum_{j=1}^{M}\sum_{i \in \mathbb{C}_{j}}\mathbb{E}\left[\dfrac{W}{M} \log \left(1+\dfrac{h_{ii}}{ \frac{N_{0}W}{M}}\right) \right], 
\end{equation}
where the expectation is computed with respect to $h_{ii}$. Under a Rayleigh fading channel 
condition and using the fact that $n=\frac{K}{M}$, (\ref{rr0}) can be written as
\begin{eqnarray}
\bar{R}_{ave} &=&  nW \int_{0}^{\infty} e^{-x}\log \left(1+\frac{M}{N_{0}W} x \right) dx \\
\label{exp04} & = & \dfrac{KW}{M} e^{\frac{N_{0}W}{M}} \mathrm{E}_{1} \left (\frac{N_{0}W}{M} \right)\\
\label{exp05}&=&\dfrac{KW}{M} e^{\frac{N_{0}W}{M}}\int_{1}^{\infty}\dfrac{e^{-t\frac{N_{0}W}{M}}}{t}dt,
\end{eqnarray}
where $\mathrm{E}_{1}(x)=-\mathrm{Ei}(-x)=\int_{1}^{\infty}\frac{e^{-tx}}{t}dt$, $x>0$. 
Taking the first-order derivative of (\ref{exp05}) in terms of $M$ yields,
\begin{equation}\label{asy}
\dfrac{\partial \bar{R}_{ave}}{\partial M}=-\dfrac{KW}{M^{2}}e^{\frac{N_{0}W}{M}} \left( 1+\dfrac{N_{0}W}{M} \right)\mathrm{E}_{1} \left (\frac{N_{0}W}{M} \right)+\dfrac{KW}{M^{2}}. 
\end{equation}
Since for every value of $N_{0}W$, $\dfrac{\partial \bar{R}_{ave}}{\partial M}$ is negative, 
it is concluded that the network average sum-rate is a monotonically decreasing function of 
$M$. Consequently, the maximum average sum-rate of the network for $\alpha = 0$ and every value 
of $1 \leq M \leq K$ is achieved at $M=1$.

\section{Proof of Remark 5}\label{append07}

\renewcommand{\theequation}{G-\arabic{equation}}
\setcounter{equation}{0}
From (\ref{eqn: 02}) and (\ref{eqn: 05}), the average sum-rate of the network is given by
\begin{eqnarray}
\bar{R}_{ave} & = & \mathbb{E}\left[\sum_{i=1}^{K} R_{i}(\hat{\textbf{\textrm{P}}}^{(j)},\boldsymbol{\mathcal{L}}_{i}^{(j)}) \right] \\
& = & \dfrac{W}{K} \sum_{i=1}^{K} \mathbb{E}\left[ \log \left(1+\dfrac{h_{ii}}{\frac{N_{0}W}{K}} \right) \right],
\end{eqnarray}
where the expectation is computed with respect to $h_{ii}$. Under a Rayleigh fading channel condition, we have 
\begin{eqnarray}
\bar{R}_{ave} & = & W \int_{0}^{\infty} e^{-x}\log \left(1+\frac{K}{N_{0}W} x\right) dx\\
\label{exp01} & = & W e^{\frac{N_{0}W}{K}} \mathrm{E}_{1} \left (\frac{N_{0}W}{K} \right).
\end{eqnarray}
To simplify (\ref{exp01}), we use the following series representation for $\mathrm{E}_{1}(x)$,  
\begin{equation} \label{Euler}
\mathrm{E}_{1}(x)= - \boldsymbol{\gamma}+ \log\left( \dfrac{1}{x}\right) +\sum_{s=1}^{\infty} \dfrac{(-1)^{s+1}x^{s}}{s.s!}, ~~x>0,
\end{equation}
where $\boldsymbol{\gamma}$ is Euler's constant and is defined by the limit \cite{Gradshteynbook94}
\begin{equation}
\boldsymbol{\gamma} \triangleq \lim_{s \to \infty} \left( \sum_{k=1}^{s} \dfrac{1}{k}-\log s \right)=0.577215665... \nonumber 
\end{equation}
Thus, (\ref{exp01}) can be simplified as
\begin{equation}
\bar{R}_{ave}=W e^{\frac{N_{0}W}{K}} \left ( - \boldsymbol{\gamma}+ \log \left(\dfrac{K}{N_{0}W} \right) +\sum_{s=1}^{\infty} \dfrac{(-1)^{s+1}}{s.s!} \left( \dfrac{N_{0}W}{K} \right)^{s} \right). 
\end{equation}
In the asymptotic case of $K \rightarrow \infty$, 
\begin{equation}
e^{\frac{N_{0}W}{K}} \approx 1,
\end{equation} 
and
\begin{equation}
\sum_{s=1}^{\infty} \dfrac{(-1)^{s+1}}{s.s!} \left( \dfrac{N_{0}W}{K} \right)^{s} \approx 0. 
\end{equation}
Consequently, the network average sum-rate for $M=K$ is asymptotically obtained by 
\begin{equation}
\bar{R}_{ave} \approx W(\log K-\log N_{0}W-\boldsymbol{\gamma}).  
\end{equation}

\end{document}